\documentclass[aps,prd,onecolumn,superscriptaddress,showpacs,nofootinbib,floatfix]{revtex4}

\usepackage{dcolumn}
\usepackage{bm}

\usepackage[dvips]{graphicx}
\usepackage{float}
\usepackage{epsfig}
\usepackage{graphicx}
\usepackage{longtable}
\usepackage{rotating}
\usepackage{amsmath}
\usepackage{ulem}
\usepackage{subfigure}
\usepackage{txfonts}

\usepackage{color}

\def\GeVc{\ensuremath{\rm{{GeV}/}c }}
\def\GeVcc{\ensuremath{\rm{{GeV}/}c^{2} }}

\def\mean#1{\ensuremath{\left<#1\right>}}
\def\ttt#1{\texttt{\small #1}}

\providecommand{\pp}{$pp$}
\providecommand{\pA}{$p\,$A}
\providecommand{\pO}{$p\,$O}
\providecommand{\pAr}{$p\,$Ar}
\providecommand{\pPb}{$p\,$Pb}
\providecommand{\AaAa}{AA}
\providecommand{\ArAr}{ArAr}
\providecommand{\OO}{OO}
\providecommand{\PbPb}{PbPb}
\providecommand{\gaga}{\gamma\,\gamma}

\newcommand{\sqrts}{\sqrt{s}}

\newcommand{\sqrtsnn}{\sqrt{s_{_{\ensuremath{\it{NN}}}}}}

\newcommand{\pPbH}{$p \,$Pb$\,\xrightarrow{\gaga} p \,H \,$Pb}
\newcommand{\pPbHX}{$p \,$Pb$\,\xrightarrow{\gaga} X \,H \,$Pb}
\newcommand{\pAH}{$p \,A\,\xrightarrow{\gaga} p \,H \,A$}
\newcommand{\pAHX}{$p \,A\,\xrightarrow{\gaga} X \,H \,A$}

\newcommand{\ABH}{$A \,B\,\xrightarrow{\gaga} A \,H \,B$}
\newcommand{\ABHX}{$A \,B\,\xrightarrow{\gaga} X \,H \,B$}

\newcommand{\ce}[1]{Eq.~(\ref{#1})}
\newcommand{\cf}[1]{{Fig.~\ref{#1}}}

\newcommand{\ttbar}    {\ensuremath{t\bar{t}}}
\newcommand{\bbar}     {\ensuremath{b\bar{b}}}
\newcommand{\QQbar}    {\ensuremath{Q\bar{Q}}}
\newcommand{\Qqbar}    {\ensuremath{\textsc q}\bar{\textsc q}}
\newcommand{\qqbar}    {\ensuremath{q\bar{q}}}
\newcommand{\ccbar}    {\ensuremath{c\bar{c}}}
\providecommand{\madgraph}{{\sc madgraph}}

\providecommand{\geant}{{\sc geant}}

\newcommand{\Lumi}{\mathcal{L}}
\newcommand{\Lunits}{cm$^{-2}$s$^{-1}$}

\newcommand{\emittn}{\epsilon_{n}}

\newcommand{\Pom} {I\!P}

\providecommand{\elel}{e^+e^-}

\begin{document}

\title{Study of Higgs boson production and its $\bbar$ decay \\[1.5 truemm] in  $\gamma$-$\gamma$ processes in proton-nucleus collisions at the LHC}

\author{David~d'Enterria}
\affiliation{ICC-UB \& ICREA, Univ. de Barcelona, 08028 Barcelona, Catalonia}

\author{Jean-Philippe Lansberg\protect\footnote{Present address at Ecole Polytechnique (Palaiseau).}}
\affiliation{Centre de Physique Th\'eorique, \'Ecole Polytechnique, CNRS,  91128 Palaiseau, France}
\affiliation{SLAC Nat. Acc. Lab., Theoretical Physics, Stanford University, Menlo Park, California 94025, USA}


\begin{abstract}
\noindent
We explore for the first time the possibilities to measure an intermediate-mass
($m_H =$~115~--~140~\GeVcc) Standard-Model Higgs boson in electromagnetic 
proton-lead (\pPb) interactions at the CERN Large Hadron Collider (LHC) via its $\bbar$ decay.
Using equivalent Weizs\"acker-Williams photon fluxes and Higgs effective field
theory for the coupling $\gaga\to H$, we obtain a leading-order cross section
of the order of 0.3~pb for exclusive Higgs production in elastic (\pPbH) 
and semielastic (\pPbHX) processes at $\sqrtsnn$~=~8.8 TeV. 
After applying various kinematics cuts to remove the main backgrounds 
($\gaga\to \bbar$ and misidentified $\gaga\to \qqbar$ events), we find that a Higgs 
boson with $m_H$~=~120~\GeVcc\ could be observed in the $\bbar$ channel with a 
3$\sigma$-significance integrating 300~pb$^{-1}$ with an upgraded \pA\ luminosity of
10$^{31}$\Lunits. 
We also provide for the first time semielastic Higgs cross sections, along with elastic 
$\ttbar$ cross sections, for electromagnetic \pp, \pA\ and \AaAa\ collisions at the LHC . \\
\noindent
\end{abstract}

\pacs{14.80.Bn, 25.20.Lj}

\maketitle


\section{Introduction}

The standard model (SM) of particle physics predicts the existence of a scalar Higgs particle 
($H$) to explain the breaking of the electroweak gauge symmetry observed 
in nature~\cite{Higgs}. Direct searches for the Higgs boson at the LEP collider 
have constrained its mass above 114.4 \GeVcc\ at 95\% confidence level (CL)~\cite{Barate:2003sz},
and global fits to precision electroweak data 
exclude $m_H > 154\ \GeVcc$~at 95\% CL~\cite{Collaboration:2008ub}. Yet, it turns out that the 
favored intermediate-mass range above the LEP limit is the most difficult region for Higgs searches 
in \pp\ collisions at the LHC. Indeed, for $m_H < 135\ \GeVcc$, the dominant 
decay mode is $H \rightarrow \bbar$ with a typical cross section 
$\sigma(H\to \bbar)\approx$~30~pb at $\sqrts$~=~14~TeV~\cite{Djouadi:2005gi}, 
which is overwhelmed by the combinatorial background from QCD $b$-jets production with $\sigma(\bbar)\approx$~500~$\mu$b.
As a matter of fact, the $H\to\bbar$ decay channel is now considered unaccessible\footnote{There are however recent 
developments~\cite{DeRoeck:2007zza,Butterworth:2008iy} that give some hope in the $W\,H(\bbar)$ and $Z\,H(\bbar)$
associated production modes.} at the LHC~\cite{atlas_tdr,cms_tdr}, 
and testing the expected mass-dependent Yukawa coupling of the Higgs boson to the $b$-quark seems to be left open 
to study only at a next $\elel$ linear collider. In the intermediate-mass range, standard Higgs searches need thus 
to resort either to rare decay modes such as $H\to \gaga$ or to very stringent cuts on the final-state particles, 
which lead to 3--4 orders of magnitude reduction of the observed cross section.\\ 

In this context, the clean topologies of {\it exclusive} Higgs production 
in ``peripheral'' \pp\ processes mediated by colorless exchanges -- such as two gluons in a color-singlet
state (Pomerons)~\cite{Khoze:2001xm,DeRoeck:2002hk} or two photons~\cite{Khoze:2001xm,Piotrzkowski:2000rx} -- 
are attracting increasing interest~\cite{d'Enterria:2008sh,louvain} despite their much smaller cross sections, 
${\cal O}(10^{-4}-10^{-5})$, compared to the dominant gluon-fusion or vector-boson-fusion (VBF) 
Higgs production channels. 
Exclusive events are characterized by wide rapidity gaps on both sides of the singly produced central
system and the survival of both protons scattered at very low angles with respect to the beam. 
The final-state is thus much cleaner with just the decaying products in the central detector, 
the signal/background is much more favorable than in parton-parton interactions, and the event 
kinematics can be constrained measuring the final protons with near-beam detectors in the LHC tunnel~\cite{fp420}.\\


In this work we consider Higgs production from $\gamma$-$\gamma$ collisions, also known as 
ultraperipheral collisions (UPCs)~\cite{Bertulani:2005ru}, with proton and ion beams at the LHC.
All charges accelerated at high energies generate electromagnetic fields which, in the equivalent 
photon approximation (EPA)~\cite{WW}, can be considered as (quasireal) photon beams\footnote{The 
emitted photons are almost on mass shell, with virtuality $- Q^{2} < 1/R^{2}$, where $R$ 
is the radius of the charge, i.e. $Q\approx$~0.28~GeV for protons ($R\approx$~0.7~fm) 
and $Q<$~0.06~GeV for nuclei ($R_A\approx 1.2\,A^{1/3}$~fm) with mass number $A>$~16.}~\cite{Stan70}. 
A significant fraction of the \pp~\cite{louvain,fp420} and \PbPb~\cite{Baltz:2007kq} 
collisions at the LHC will involve for the first time $\gamma$-induced interactions at TeV energies with 
effective photon luminosities never reached before. 
The highest available photon energies are of the order of the inverse Lorentz contracted
radius $R$ of the source charge, $\omega_{max}\approx\gamma/R$.
The photon spectrum is thus harder for smaller charges, 
which favors proton 
over nuclear beams in the production of heavy particles. However, since the photon flux scales as 
the squared charge of the beam, $Z^2$, two-photon cross sections 
are extremely enhanced for ion beams ($Z_{Pb}^4$~=~5$\cdot$10$^{7}$ for lead-lead). 
Particle production in two-photon interactions at hadronic colliders has been studied at 
RHIC by PHENIX~\cite{Afanasiev:2009hy} and STAR~\cite{STARrho,Adams:2004rz}, and at 
the Tevatron by CDF~\cite{Abulencia:2006nb,Aaltonen:2009cj,Aaltonen:2009kg}. 
They are also part of the ALICE~\cite{alice-ppr1,alice-ppr}, ATLAS~\cite{Kepka:2008zza,Pozdnyakov:2008zz}, 
CMS~\cite{cms_totem,cms_hi_ptdr} and LHCb~\cite{lhcb,lhcb_gaga} physics programmes with proton 
and/or nuclear beams. 
Two-photon fusion favors the production of spinless resonances ($\gaga\to\,$vector is forbidden by the Landau-Yang theorem).
In this work, we propose to exploit the photon fluxes generated by both the proton 
and Pb ions at the LHC to study the possible production of the SM Higgs boson. Higgs photon-fusion production 
in proton or nucleus collisions at TeV energies is not a new idea (see e.g.~\cite{Baur:2001jj} for references). 
Pioneering calculations for \pp~\cite{Papageorgiu:1995eg} and \PbPb~\cite{Grabiak:1987uf,Papageorgiu:1988yg,Drees:1989vq} 
collisions, updated more recently~\cite{Khoze:2001xm,Miller:2007pc,Baltz:1997di,Kryukov:2008zz}, 
predict cross sections of a SM Higgs boson with $m_H$~=~120~\GeVcc\ in the $\sigma_{pp\to\gaga\to H}$~=~0.1~fb 
and $\sigma_{PbPb\to\gaga\to H}$~=~10~pb ranges respectively. Unfortunately, the small value of the 
\pp\ cross section and the large event pileup\footnote{{\it Pile-up} refers to the overlapping events occurring in the 
same bunch-crossing at high luminosity $\Lumi$. For the nominal \pp\ running at $\Lumi$~=~10$^{34}$~cm$^{-2}$s$^{-1}$ 
one expects about 25 simultaneous collisions at the LHC.} in \pp\ collisions on the one hand,
and the very low design luminosities expected for \PbPb, on the other, preclude any real measurement of the 
Higgs boson in those channels.\\

Our proposal to study $\gaga \to H$ in  \pA\ collisions presents advantages with respect to both 
ultraperipheral \AaAa\ and \pp\ collisions. First, compared to \AaAa, one benefits from (i) beam 
luminosities more than 4 orders of magnitude larger: ${\cal L}_{pPb} \sim 10^{31}$~\Lunits
versus ${\cal L}_{\rm PbPb} \sim 10^{27}$~\Lunits, (ii) higher beam-beam c.m. energies: 
$\sqrtsnn$~=~8.8 TeV compared to 5.5 TeV, (iii) higher $\gaga$ c.m. energies (harder proton photon  
spectrum and smaller distance of approach between the centers of the radiating charges), and (v) easy removal of other 
photoproduction backgrounds characterized in the \AaAa\ case by additional photon exchanges 
which lead to forward neutron(s) emission. 
The net result is that one can reach higher masses and yields for any centrally produced system $H$. 
The advantage with respect to \pp\ UPCs is threefold: (i) a $Z^2$ increase in one of the photon fluxes, and
the possibilities (ii) to trigger on and carry out the measurement with almost no pileup, 
and (iii) to remove most of the exclusive diffractive backgrounds -- since the nucleus is a fragile object, 
Pomeron-mediated interactions in \pA\ will, at variance with \pp, almost always lead to the emission of a few nucleons detectable in the zero degree calorimeters.\\

\begin{figure}[hbt!]
  \centering
  \subfigure[Elastic case]{\includegraphics[height=4cm]{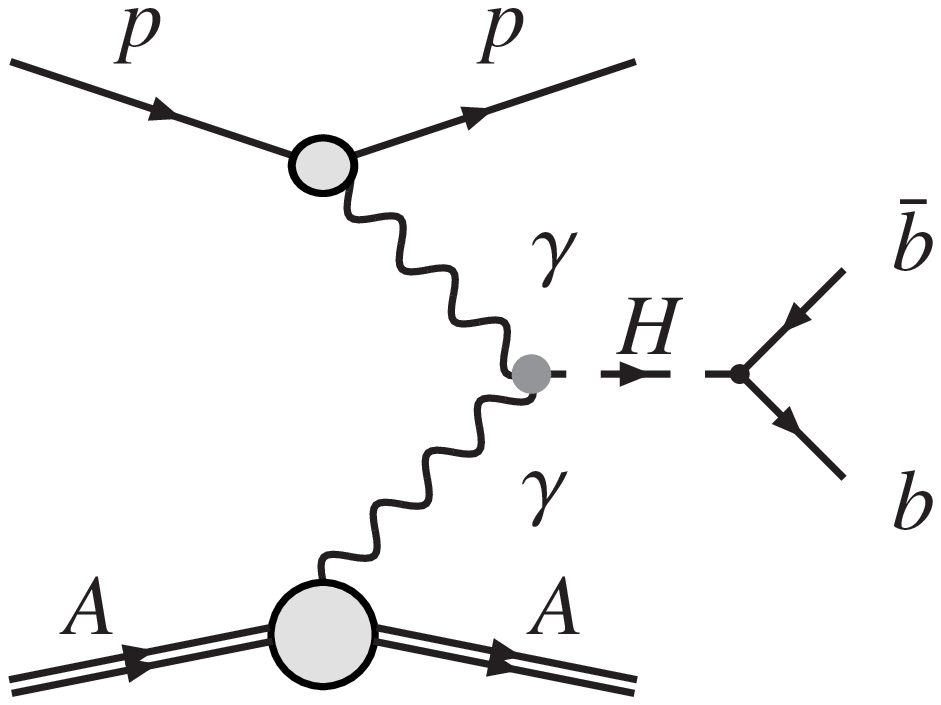}\hspace{1.cm}}
  \subfigure[Semielastic case]{\includegraphics[height=4cm]{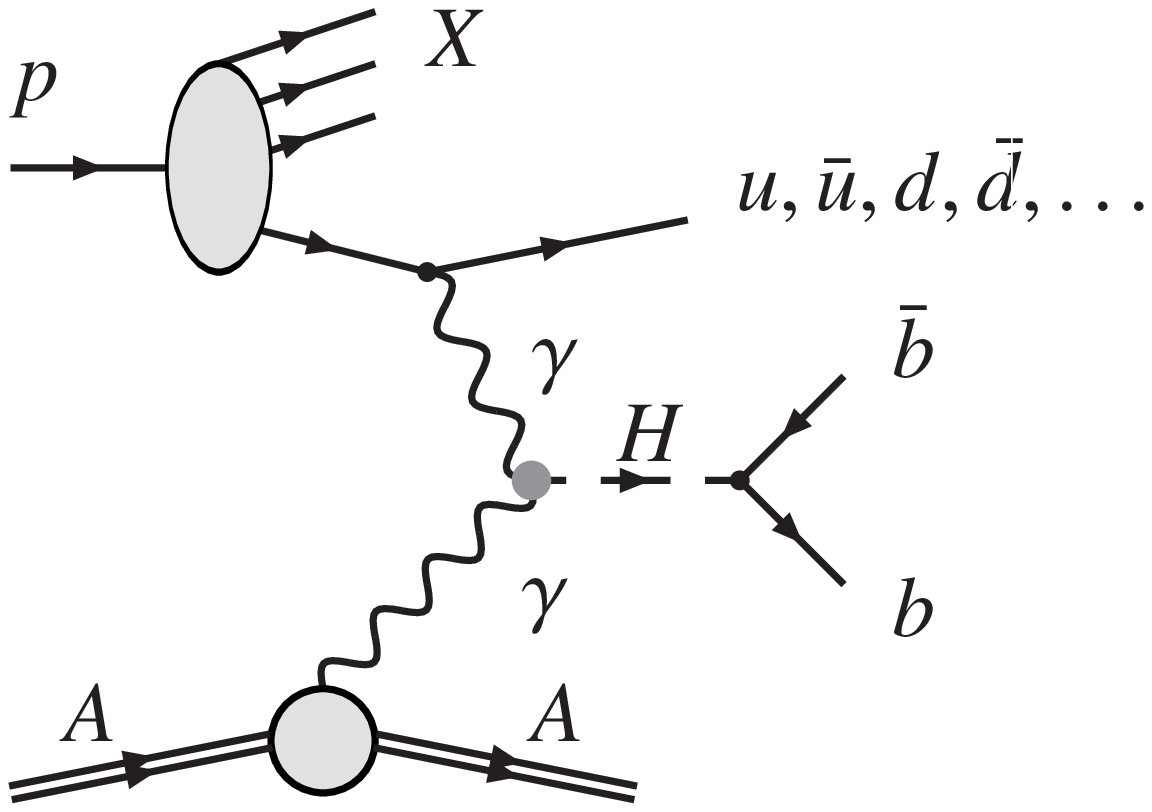}}
  \caption{Feynman diagrams for two-photon collisions with proton and nucleus beams producing a Higgs boson 
decaying into $H\to \bbar$. (a) Elastic production (both photons are emitted coherently and the proton and nucleus survive).
(b) Semielastic production (the photon from the proton is emitted by a quark, the proton subsequently breaks up).}
  \label{fig:diags}
\end{figure}

In the following, we present a detailed generator-level study for the exclusive \pPbH\ 
(elastic) and \pPbHX\ (semielastic) processes (Fig.~\ref{fig:diags}), obtained 
with the \madgraph\ code supplemented with nuclear equivalent photon spectra.
We compute the tree-level SM cross sections for the signal -- in the Higgs effective field theory (HEFT) approximation --
and for the expected backgrounds. We determine the expected yields in 1-year run taking into account the maximum 
attainable \pPb\ luminosities. We then discuss the reconstruction of the $H\to\bbar$ decay for a Higgs boson with 
$m_H$~=~120~\GeVcc, including the trigger and analysis cuts needed to minimize the backgrounds. After accounting for basic 
detector reconstruction effects ($b$-jet misidentification, $\bbar$-jets invariant mass resolution), we determine the expected statistical
significance of the measurement. Our results are promising in various fronts. First, they indicate that the study 
of the difficult $H$--$b$-quark coupling could be accessible in this production mode at the LHC. 
Second, the observation of the $\gaga\to H$ process provides an independent measurement of the Higgs-$\gamma$ coupling 
(likely measured previously in the traditional $H\to \gaga$ discovery channel). The $\gaga$-Higgs cross section is generated at the
one-loop level by all heavy charged particles ($W$ and top-quark in the SM) and is thus sensitive to possible contributions of
new charged particles with masses beyond the energy covered directly by the collider: e.g. via chargino and top-squark loops in 
supersymmetric (SUSY) extensions of the SM.

\section{Cross section evaluation}
\label{sec:sigma_calc}

The cross section for a two quasireal photon process in hadronic collisions to produce a final state $H$ at center-of-mass 
(c.m.) energy $W_{\gaga}$ (e.g. a particle $H$ with mass $m_H$, see Fig.~\ref{fig:diags}), factorizes into 
the product of the elementary cross section for $\gaga\rightarrow H$ convoluted with the equivalent photon spectra 
from the two colliding beams:
\begin{equation}
\sigma(A\;B\,\xrightarrow{\gaga} A \; H \; B)=
\int d\omega_1 d\omega_2 \, \frac{f_{\gamma/A}(\omega_1)}{\omega_1}\, \frac{f_{\gamma/B}(\omega_2)}{\omega_2} \; \sigma(\gamma\gamma\rightarrow H(W_{\gaga})),
\label{eq:two-photon}
\end{equation}
where $\omega_1$ and $\omega_2$ are the two photon energies, and $f_{A,B}(\omega)$ are the photon fluxes 
at energy $\omega$ emitted by the hadrons $A$ and $B$. 
The photon energies determine the c.m. energy $W_{\gaga}=\sqrt{s_{\gaga}}=\sqrt{4\omega_1\omega_2}$ 
and the rapidity $y$ of the produced system:
\begin{equation}
\omega_{1,2} = \frac{W_{\gaga}}{2} e^{\pm y}, \;\;\mbox{ and }\;\; y = 0.5\, \ln(\omega_1/\omega_2).
\end{equation}
For symmetric systems, the maximum effective two-photon energy $W_{\gaga}^{max}$ occurs at $y=0$, when
$\omega_1^{max}=\omega_2^{max} \approx \gamma/b_{min}$ where $\gamma=\sqrtsnn/(2m_N)$ 
is the Lorentz relativistic factor\footnote{$m_N$~=~0.9315~\GeVcc\ for nuclei, and $m_p$~=~0.9383~\GeVcc\ for protons.} 
and $b_{min}\approx 2\,R_A$ the minimum separation between the two charges of radius $R_A$.
Note that these ``maximum''  photon energies do not have to be interpreted as a hard cut-off but as an
indication of the energy ($\omega> \gamma\,\beta/b$) from which the photon flux is exponentially suppressed.\\ 

Table~\ref{tab:1} summarizes the most relevant parameters for ultraperipheral \pp, \pA, and \AaAa\ 
collisions at the LHC~\cite{Baltz:2007kq,pAyellowreport}.
It has to be noted that whereas proton-proton and nucleus-nucleus collisions are obviously part of the 
approved baseline physics programme of the LHC, proton-nucleus collisions are still considered as 
an upgrade of the heavy-ion programme~\cite{pAyellowreport,Jowett:2006au}. Detailed studies~\cite{Jowett:2006au} 
have nonetheless shown that is a perfectly feasible mode of operation of the collider, and its physics 
possibilities have been discussed vastly in the literature as a crucial baseline for the interpretation of 
the \AaAa\ data~\cite{pAyellowreport,Abreu:2007kv,Accardi:2004be}. At the LHC, protons and ions have to
travel in the same magnetic lattice\footnote{The magnetic rigidity is defined as $p/Z = B\,r$ for an ion
with momentum $p$ and charge $Z$ that would have a bending radius $r$ in a magnetic field $B$.} 
i.e. the two beams have to have the same charge-to-mass ratio $Z/A$. 
This limits the beam momentum of a given species to $p = 7 \, {\rm TeV} \, Z/A$ for the nominal 8.3~T
dipole bending field. Thus, the energy in the nucleon-nucleon center-of-mass is 
$\sqrtsnn = 14 \, {\rm TeV} \, \sqrt{(Z_{A} Z_{B})/(A \, B)}$. In the case of $AB$ collisions with 
asymmetric beam energies the rapidity of the c.m. system shifts relative to the laboratory by
$\Delta y_{\rm c.m.} = 0.5 \, \ln{[(Z_{A} \, B )/(Z_{B} \, A )]}$. Thus, for \pA\ collisions the
rapidity shifts span a range of $\Delta y_{\rm c.m.}$~=~0.35 -- 0.47 for ions from oxygen to lead. 
For $\gaga$ processes in ultraperipheral \pA\ collisions, the average rapidity shift is even larger due to the 
harder EPA spectrum from the proton (e.g.  $\langle \Delta y\rangle \approx$~1.7 for 
the case of 120 \GeVcc\ Higgs boson, see Section~\ref{sec:Hbbar}). 
The quoted \pA\ luminosities have to be considered as unofficial but plausible values 
(see Section~\ref{sec:rates_lumi}). The Table indicates that the ``maximum'' $\gaga$ c.m. energies 
attainable range from $\sqrt{s_{\gaga}^{\rm max}}\approx$~160 GeV for \PbPb\ to 4.5~TeV for \pp\
(obtained taking $R_p$~=~0.7~fm). Two-photon fusion collisions in \pPb\ have
$\sqrt{s_{\gaga}^{\rm max}}\approx$~260~GeV, i.e. more than twice the most probable mass of the SM Higgs boson.

\begin{table}[htpb]
\begin{center}
\caption[]{Relevant parameters for photon-induced processes in $A\,B$ collisions at the LHC:
(i) beam luminosity, $\Lumi_{AB}$, (ii) nucleon-nucleon c.m. energy, $\sqrtsnn$, 
(iii) beam energies, $E_{\rm beam}$, (iv) Lorentz factor, $\gamma=\sqrtsnn/(2\,m_N)$,  (v) effective radius 
(of the largest species), $R_A$, (vi) photon ``cutoff energy'' in the c.m. frame, $\omega_{\rm max}$, 
(vii) ``maximum'' photon-nucleon c.m. energy, $\sqrt{s_{\gamma_N}^{\rm max}}=\sqrt{2\omega_{\rm max}m_N}$, 
(viii) ``maximum'' photon-photon c.m. energy, $\sqrt{s_{\gaga}^{\rm max}}$, and (ix) hadronic cross section, $\sigma_{inel}$ 
(the \pp\ value is from~\cite{Cudell:2002xe}, the \pA\ and \AaAa\ geometric cross 
sections are obtained with a Glauber model with $\sigma_{_{inel,NN}}\approx$~80~mb~\cite{d'Enterria:2003qs}).}
\label{tab:1}
\vspace{0.4cm}
\begin{tabular}{|l|c|c|c|c|c|c|c|c|c|} \hline\hline
System  & $\sqrt{s_{_{NN}}}$ & ${\cal L}_{AB}$  & $E_{\rm beam1}$ + $E_{\rm beam2}$ & $\gamma$ &  $R_A$ & $\omega_{\rm max}$ 
      & $\sqrt{s_{\gamma_N}^{\rm max}}$ & $\sqrt{s_{\gaga}^{\rm max}}$ & $\sigma_{inel}$ \\
       & (TeV) & (cm$^{-2}$s$^{-1}$) & (TeV) &  & (fm) & (GeV) & (GeV) & (GeV)  & (mb) \\ \hline
\pp  & 14 & $10^{34}$   & 7. + 7. & 7455 & 0.7 & 2450  & 8400 & 4500 & 110 \\ \hline 
\pO & 9.9 & $2.7\cdot 10^{30}$ & 7. + 3.5 & 5270 & 3.0  & 340  & 2600 & 690 & 480 \\ 
\pAr & 9.4 & $1.5\cdot 10^{30}$ & 7. + 3.15 & 5000 & 4.1 & 240  & 2130 & 480 & 830 \\  
\pPb & 8.8 & $1.5\cdot 10^{29}$ & 7. + 2.76 & 4690 & 7.1 & 130  & 1500 & 260 & 2160 \\ \hline 
\OO   & 7.0 & $2.\cdot 10^{29}$ & 3.5 +3.5 & 3730 & 3.0 &  240 & 1850 & 490 & 1500 \\ 
\ArAr & 6.3 & $0.6\cdot 10^{29}$  & 3.15 + 3.15 & 3360 & 4.1 & 160 & 1430 & 320 & 2800 \\ 
\PbPb & 5.5 & $5\cdot 10^{26}$ & 2.76 + 2.76 & 2930 & 7.1 &  80  &  950 & 160 & 7700 \\ \hline\hline
\end{tabular}
\end{center}
\end{table}

\subsection{Equivalent photon fluxes}

In the Weizs\"acker-Williams approximation~\cite{WW}, the flux of equivalent photons from a 
relativistic particle of charge $Z$ is determined from the Fourier transform of its electromagnetic field.
For an extended charge with form-factor $F(Q^2)$, such as a proton or a nucleus, the energy spectrum 
$f_{\gamma/A}(x) = dn_{\gamma}/dx$, where $x = \omega/E$ is the fraction of the beam energy 
carried by the photon, can be calculated from~\cite{Budnev:1974de}:
\begin{equation}
f_{\gamma/A}(x) = \frac{\alpha Z^2}{\pi} \, \frac{1 - x + 1/2 x^2}{x} 
\int_{Q_{min}^2}^{\infty} \frac{Q^2 - Q_{min}^2}{Q^4} | F(Q^2) |^2 dQ^2 \;,
\label{eq:f_x}
\end{equation}
where $\alpha=1/137$, and $Q^2$ is the 4-momentum transfer squared from the charge.
The minimum momentum transfer squared, $Q_{min}^2 \approx (x m_A)^2/(1-x)$, is a function 
of $x$ and the mass $m_A$ of the projectile.\\

For UPCs involving ions it is more appropriate to calculate the spectrum of equivalent photons 
as a function of impact parameter~\cite{Cahn:1990jk,Baur:1990fx}. 
The photon energy spectrum produced by a charge $Z$ sweeping past a target, integrated on the impact parameter
$b$ from $b_{min}$ to infinity, is a textbook analytical result~\cite{Jackson}:
\begin{equation}
f_{\gamma/A}(x) = \frac{\alpha Z^2}{\pi} \, \frac{1}{x} \, \bigg[ 2 x_i K_0(x_i) K_1(x_i) - x_i^2 (K_1^2(x_i) - K_0^2(x_i)) \bigg] \; ,
\label{eq:flux_A}
\end{equation}
where $x_i= x\, m_N \, b_{min}$, and $K_0$, $K_1$ are the modified Bessel functions of the second kind
of zero and first order, related respectively to the emission of longitudinally and transversely polarized photons. 
The transverse polarization dominates for ultrarelativistic particles ($\gamma\gg1$). 
Although this approach treats the nucleus as an idealistic hard sphere, the use of more realistic Woods-Saxon profiles
gives effective $\gaga$ luminosities only about 5\% lower~\cite{Baltz:2009jk} than the hard-sphere approximation 
in the range of c.m. energies $W_{\gaga}\approx 0.5\,W_{\gaga}^{max}$ dominant in the exclusive production 
of an intermediate-mass Higgs boson.\\

\begin{figure}[tbhp]
\centering
\includegraphics[width=12.cm]{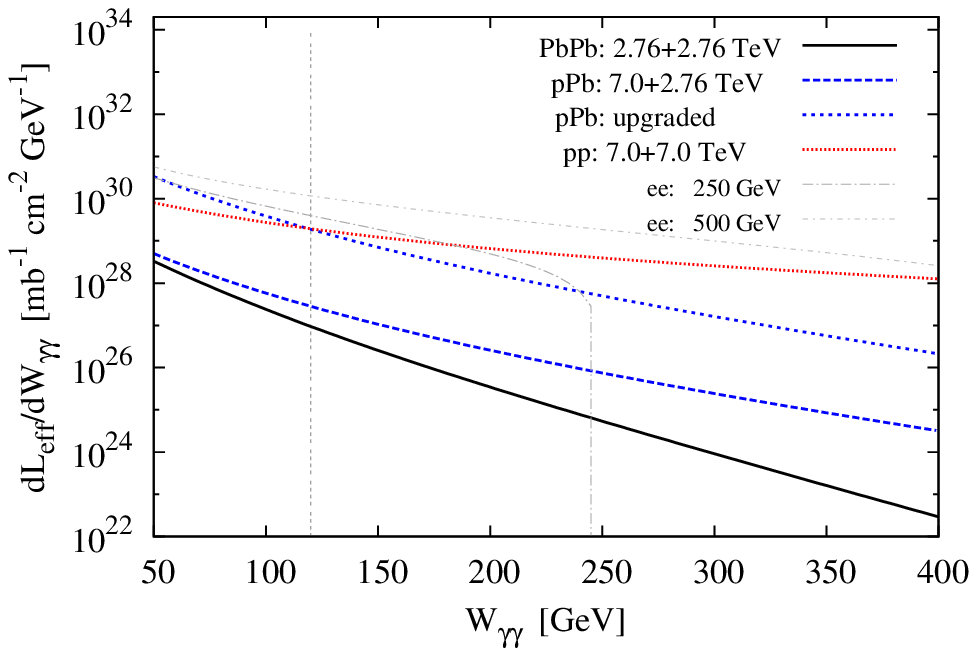}
  \caption{
Effective elastic two-photon luminosities ${\cal L}_{AB}(d{\cal L}_{\gamma\gamma}/dW_{\gaga})$
for \pp, \pPb\ and \PbPb\ collisions at the LHC based on 
the beam luminosities quoted in Table~\ref{tab:1}. For \pPb\ we show also the curve corresponding to a 
$\times$60 luminosity upgrade (Sect.~\ref{sec:rates_lumi}). The effective $\gaga$ luminosities in $e^+e^-$
collisions at the ILC are also shown for $\sqrts$~=~250 GeV and 500 GeV (thin dashed lines)~\cite{Pire:2008xe}.}
  \label{fig:lumis}
\end{figure}

The figure of merit for $\gaga$ processes in UPCs is 
$\Lumi_{\gaga}^{\textrm{eff}} \equiv \Lumi_{AB}\,d\Lumi_{\gaga}/dW_{\gaga}$ where 
${\cal L}_{AB}$ is the collider luminosity for a given $A\,B$ system and $d\Lumi_{\gaga}/dW_{\gaga}$ 
is the photon-photon luminosity as a function of the $\gaga$ c.m. energy obtained integrating the 
two photon fluxes over all rapidities $y$,
$d^2\Lumi_{\gaga}/dW_{\gaga}dy=(2/W_{\gaga})f_{\gamma/A}(W_{\gaga}/2 e^{y})f_{\gamma/B}(W_{\gaga}/2 e^{-y})$. 
For illustration, in Fig.~\ref{fig:lumis} we show $\Lumi_{\gaga}^{\textrm{eff}}$ obtained 
from the parametrization of ref.~\cite{Cahn:1990jk} of $d\Lumi_{\gaga}/dW_{\gaga}$ for ion-ion collisions
and using the $\Lumi_{AB}$ luminosities quoted\footnote{We note, as pointed out in~\cite{Pire:2008xe}, 
that the $\Lumi_{\gaga}^{\textrm{eff}}$ plot (Fig.~3) of~\cite{Baltz:2007kq}, does not use correct (updated) values for $\Lumi_{pp}$ at the LHC.} 
in Table~\ref{tab:1}.  The curves are computed for the elastic $\gaga$ fluxes. Inclusion of the semielastic fluxes 
would yield luminosities twice higher, as we discuss later. For comparison, we also plot the effective $\gaga$ 
luminosities in $e^+e^-$ collisions at the ILC (${\cal L}_{e+e-}$~=~2$\cdot$10$^{34}$~\Lunits) 
for $\sqrts$~=~250 GeV and 500 GeV~\cite{Pire:2008xe}.\\

The elastic \pPb\ two-photon luminosities at the LHC are similar to those for \PbPb\ for low $\gaga$ center 
of mass energy, $W_{\gaga}$, and become higher for $W_{\gaga}>50$~GeV due to the larger \pPb\ beam 
energies. For energies of interest for an intermediate-mass Higgs (dotted vertical line in Fig.~\ref{fig:lumis}), 
both $\gaga$ luminosities are still almost two order of magnitude lower than that in proton-proton collisions. However, 
as we discuss in Section~\ref{sec:rates_lumi}, there are seemingly no technical reasons that would prevent 
one to increase the instantaneous proton-nucleus luminosity by up to a factor $\mathcal{O}$(60) (third curve
in Fig.~\ref{fig:lumis}). We remark that our study, based on \madgraph, does not make direct use of the 
effective two-photon luminosities plotted in Fig.~\ref{fig:lumis}, 
although it gives similar results for the fluxes as we explain below.

\subsubsection{Elastic production (\pAH):}
\label{sec:elastic}

In the case of high-energy protons, the equivalent 
photon spectrum 
can be obtained from its elastic form factors in the dipole approximation, $F_M=G_M^2$ and 
$F_E=(4m_p^2G_E^ 2+Q^2G_M^2)/(4m_p^2+Q^2)$ 
with $G_E^2 = G_M^2 /\mu_p^2 = (1+Q^2/Q_0^2)^{-4}$, where 
$Q_0^2\approx0.71\,{\rm GeV}^2$ and $\mu_p^2\approx 7.78$, and reads~\cite{Budnev:1974de}:
\begin{eqnarray}
f_{\gamma/p}(x) &=& \frac{dn_{\gamma}}{dx} = \frac{\alpha}{\pi}\,\frac{1-x}{x}\;\left[\varphi(x,Q^2_{max}/Q_0^2)-\varphi(x,Q^2_{min}/Q_0^2)\right]\;,
\mbox{where} \label{eq:flux_p}\\
 \varphi(x,Q)  &= & (1+a\,y)\,\Big[-\ln\,\tfrac{1+Q}{Q}+\sum_{k=1}^3\tfrac{1}{k\,(1+Q)^k}\Big] \, + \, \tfrac{(1-b)\,y}{4Q(1+Q)^3}\,
+ c\,\left(1+\tfrac{y}{4}\right)\,\Big[\ln\,\tfrac{(1+Q)-b}{1+Q}+\sum_{k=1}^3\tfrac{b^k}{k(1+Q)^k}\Big]
\end{eqnarray}
with $y=x^2/(1-x)$. The parameters $a$, $b$ and $c$ are given by
$a=(1+\mu_p^2)/4+4\,m_p^2/Q_0^2\approx$~7.16,  $b=1-4m_p^2/Q_0^2\approx$~-3.96,
and $c=(\mu_p^2-1)/b^4\approx$~0.028. This flux\footnote{The precise value of $Q^2_{max}$ does 
not actually matter since the flux is already negligible for $Q^2$ larger than 2 GeV$^2$.}, with 
$Q^2_{max} \approx \mu_F^2$, is implemented in the standard version of  
\madgraph\ v.4~\cite{madgraph}. The use of more rigorous expressions e.g. including the magnetic 
dipole moment of the proton~\cite{Kniehl:1990iv} results only on small differences in the final 
EPA fluxes~\cite{Nystrand:2004vn}.\\


In UPCs involving nuclei one could imagine, starting from~\ce{eq:flux_A}, to compute the two photons flux by simple 
multiplication and identifying $b_{min}$ with the radii of the two nuclei $A$ and $B$ for each for photon flux. However, 
we would then miss geometrical constraints in order to have the two photons produced at the same point, outside the nuclei 
while the nuclear ``halos" do not overlap. In a $AB$ collision this implies requiring not only $b_1 > R_A$, $b_2 > R_B$, but also
$|\vec b_1 - \vec b_2| > R_A + R_B$. The latter condition, which excludes any overlapping configurations, prevents 
to derive a factorized formula for the $\gaga$ flux in terms of the momentum fraction $x_1$ and $x_2$ carried by each 
photon from their parents nuclei in~\ce{eq:flux_A}. Photon fluxes are encoded in \madgraph\ in 
the same way as any parton distribution and one has to use factorized expressions.  One option to overcome those 
complications, as proposed by Cahn~\cite{Cahn-proc}, would be to impose $b_1 > R_A+R_B$, $b_2 > R_A+R_B$ 
which together always exclude overlaps. However, this leads to quite pessimistic estimates of the joint flux~\cite{Cahn:1990jk}
since it prevents configurations where the two nucleus are very close and produce very energetic $\gaga$ collisions.
An alternative is simply not to exclude overlaps and to impose $b_1 > R_A$ and $b_2 > R_B$ alone, i.e. to use \ce{eq:flux_A} for both fluxes.
As discussed in~\cite{Cahn:1990jk}, in such a case, the deviation from an exact evaluation of the flux basically depends on the ratio
of the invariant mass of the $\gaga$ system over the c.m. energy. For the production of a Higgs with $m_H$~=~120~\GeVcc~
in \PbPb\ collisions at the LHC, one expects a deviation of a factor of 2. This is by far the most extreme case since the Pb
nuclei are large and the available energy  is reduced forcing the $\gaga$ creation point to be very near the nuclei, including
overlapping configurations. For \ArAr\ and \OO\ collisions, the corrections are already much smaller~\cite{Cahn:1990jk}. 
A comparison of the cross sections which we obtain with factorized fluxes without full non-overlap condition (see later 
Table~\ref{tab:sigma_signal}, third column) with the results for those systems using the exact non-overlap 
condition~\cite{Kryukov:2008zz} confirms the small impact of the approximations used here.\\

In the $pA$ case, using Eq.~(\ref{eq:flux_A}) with $b_2 > R_A$ and Eq.~(\ref{eq:flux_p}) 
is in fact not problematic since (i) the available energy is higher than in \AaAa\ UPCs, 
(ii) the overlap between the proton and the nucleus $A$ is reduced by the small size of the proton 
($R_p\approx$~0.7~fm)\footnote{Of course, the proton is not a sphere and its radius is not a well defined quantity. 
Effectively, electron-proton scattering fits yield a charge rms-radius $R_p$~=~0.89~$\pm$~0.2~fm~\cite{Sick:2003gm}, 
and diffractive results at HERA indicate an effective proton transverse size 
$R_p$~=~0.65~$\pm$~0.2~fm~\cite{Diaconu:2007zza}.}, and (iii) on 
average, the photon radiated by the proton (for which $b_{min}\simeq 0.7$ fm) is much more energetic\footnote{Hence
the larger shift in the rapidity distribution of a particle $X$ in $p\,A\,\xrightarrow{\gaga}X$ compared to 
hadronic $p\,A\,\to X$ collisions (see later).}. As a consequence, the $\gaga$ production point
is typically far from the nucleus surface and the proton cannot overlap with the latter. Quantitatively, we have found
that the photon momentum fraction from a Pb nucleus in a \pPb\ collisions is typically below 0.015, corresponding to
$b_2 \gtrsim 2 R_{Pb}$.
Therefore, in our calculations, we have used the equivalent photon spectrum given by Eq.~(\ref{eq:flux_p}) for protons
and by Eq.~(\ref{eq:flux_A}) for ions with the requirement $b_{min}=R_A$, where the effective 
nuclear radii are obtained in the standard way from their mass number $A$ 
via $R_A$~=$r_0\,A^{1/3}$ with $r_0$~=~1.2~fm. 
As discussed above, in practice this condition, along with the proton flux, 
ensures that the final state is produced {\it exclusively} 
and outside of the colliding system \pA, i.e. it avoids the hadronic overlap and breakup of the colliding beams.
The fact that our $\gaga$ cross sections agree well with other recent 
calculations~\cite{Khoze:2001xm,Kryukov:2008zz} (see later) lends support to our 
approximations.\\

\subsubsection{Semielastic production (\pAHX):} 
\label{sec:semiel}

In semielastic production, \cf{fig:diags}(b), the proton does not radiate coherently. The photon flux is generated 
by its quarks and is followed by the proton breakup. As found previously~\cite{Drees:1994zx,Ohnemus:1994xf,Bhattacharya:1995id} 
for two-photon fusion processes in \pp\ collisions, we expect in the $pA$ case a comparable magnitude of the incoherent 
photon flux ($\propto A=1$) emitted by the quarks compared to the flux from coherent elastic emission ($\propto Z^2=1$).
The same is not true for UPCs involving two nuclei where the elastic contribution clearly dominates over the semielastic one 
(either from the constituent protons or quarks of the nuclei) due to the large $Z^2$ factor in both coherent fluxes. 
For semielastic \pAHX\ collisions, one can consider the proton flux as a partonic distribution $\gamma^p(x,Q^2)$, the photon 
being the parton, where $x$ denotes the momentum fraction of the ``inelastic'' photon in the proton and $Q^2$ the resolution 
scale at which the proton is probed. It can be approximated by~\cite{Drees:1994zx,Ohnemus:1993qw,Ohnemus:1994xf}
\begin{equation}
\gamma^p(x,Q^2)=\frac{\alpha}{2\pi} \log\frac{Q^2}{Q^2_0} \sum_q \int_x^1 \frac{dy}{y} P_{\gamma q}(x/y) \;\left[q(y,Q^2)+\bar q(y,Q^2)\right]
\label{eq:flux_p_inel}
\end{equation}
with $P_{\gamma q}(z)=e^2_q \,(1+(1-z)^2)/z$, $Q^2_0$ an energy cut-off, and $q(x,Q^2)$ the quark PDFs in the proton.
An improved expression was further discussed in~\cite{Gluck:1994vy}.
In our work, we shall take advantage of  \madgraph\ features and generate the semielastic contribution by 
considering the partonic processes $q\gamma \to q \,H\to q \,b \bar b$ (where $q=u,\bar u,d,\bar d, s, \bar s$ 
are all possible radiating light-quarks) convoluted with the quark PDFs in the proton\footnote{We have used 
CTEQ6L.1~\cite{Pumplin:2002vw}.} and the photon flux from the nucleus $A$. 
The contributions from all the light quarks but the $u$ are found to amount to 1/3 of the $u$-quark 
alone, since they are comparatively suppressed by their smaller PDFs in the proton and/or by their lower quark-charge.\\

Obviously, one has to introduce an effective mass for the quark otherwise the cross section would be logarithmically 
divergent as \ce{eq:flux_p_inel} is if $Q^2_0$ is set to 0. We take this effective mass to be $\widehat m_q=300$ MeV. 
In order to limit the off-shellness of the quasireal photon, we have also found reasonable to bound the maximum transverse 
momentum of the outgoing quark to $p_{T,max}^{q'}=5$ GeV/c. This provides us at the same time with a natural value 
for the factorization scale entering the PDF for the incoming quark, $Q^2=(p^{q'}_{T,max})^2$.\\

In the semielastic class of events, hadrons from the fragmentation of the radiating quark and from the proton remnants spray in the 
proton-direction hemisphere, while a pure $\gamma$ beam, unaccompanied by hadronic activity, is generated by the nucleus in the opposite 
hemisphere. To guarantee a wide enough rapidity-gap as expected in exclusive production, we exclude events where the jet 
initiated by the radiating quark $q'$ ends up in the central region. All our semielastic cross sections are thus computed with the 
condition $|y_{q'}|>$~2.5.\\

\subsection{ $\gaga\to H$ coupling (Higgs Effective field theory)} 

The coupling of the scalar Higgs to photons is mostly mediated by $W$- and top-quark loops. 
The Higgs effective field theory (HEFT) model~\cite{heft}, where the Higgs boson couples 
directly to photons, can be used as an approximation of the Standard Model. For a not too heavy  
($m_H<2m_t$) and not very energetic ($p_T<2m_t$) Higgs, it is a good approximation to take the 
mass of the heavy quark in the loop to infinity. In the limit of small Higgs masses -- below about 
$m_H$~=~150~\GeVcc\ which satisfies $m_H^2/(4m_W^2)\lesssim$~1 -- the loop induced 
interaction can be approximately described by the Lagrangian
\begin{equation}
{\cal L}_{\gaga\,H}^{\textrm{eff}}=-\frac{1}{4}g\;F_{\mu\nu} \,F^{\mu\nu} \,H\;,
\label{eq:heft}
\end{equation}
where $F_{\mu\nu} = \partial_\mu A_\nu - \partial_\nu A_\mu$ is the photon field strength tensor.
Due to the Abelian nature of QED there is only one effective vertex $g$ between photons and Higgs bosons. 
The value for the coupling constant in the HEFT model as implemented in \madgraph\ is given by
\[g=-\frac{\alpha}{\pi v} \frac{47}{18}\Big( 1+ \frac{66}{235}\tau_w +\frac{228 }{1645}\tau_w^2+
\frac{696}{8225}\tau_w^3+\frac{5248}{90475} \tau_w^4+\frac{1280}{29939}\tau_w^5+
\frac{54528}{1646645}\tau_w^6-\frac{56}{705} \tau_t-\frac{32}{987}\tau_t^2\Big),\]
where $\tau_t=m_H^2/(4m_t^2)$ and $\tau_w=m_H^2/(4m_W^2)$. 
Higher order $\tau_t$ and $\tau_w$ terms have been neglected.


\section{Results I: Cross sections and rates} 

We employ the \madgraph\ v.4 Monte Carlo~\cite{madgraph} with the elastic and semielastic proton photon 
fluxes discussed in the previous Section 
together with the nucleus photon flux Eq.~(\ref{eq:flux_A}) and the HEFT model for the Higgs-photon coupling, 
Eq.~(\ref{eq:heft}), to compute the Higgs boson cross sections in  two-photon fusion processes for the systems of Table~\ref{tab:1}. 
The Higgs decay branching ratio to $\bbar$ is obtained in \madgraph\ with {\sc hdecay}~\cite{Djouadi:1997yw},
e.g. BR$(H\to\bbar)\approx$~72\% for $m_H$~=~120~\GeVcc.
We compute also the SM cross sections for the exclusive production of 
$\bbar$ and (possibly misidentified) $\ccbar$ and light-quark ($u,d,s$) pairs, which 
constitute the most important physical background for the measurement of the $H\to \bbar$ channel.

\subsection{Signal cross sections: $\gaga\to H$} 

\begin{figure}[htbp]
\centering
\includegraphics[width=0.62\columnwidth,height=9.cm]{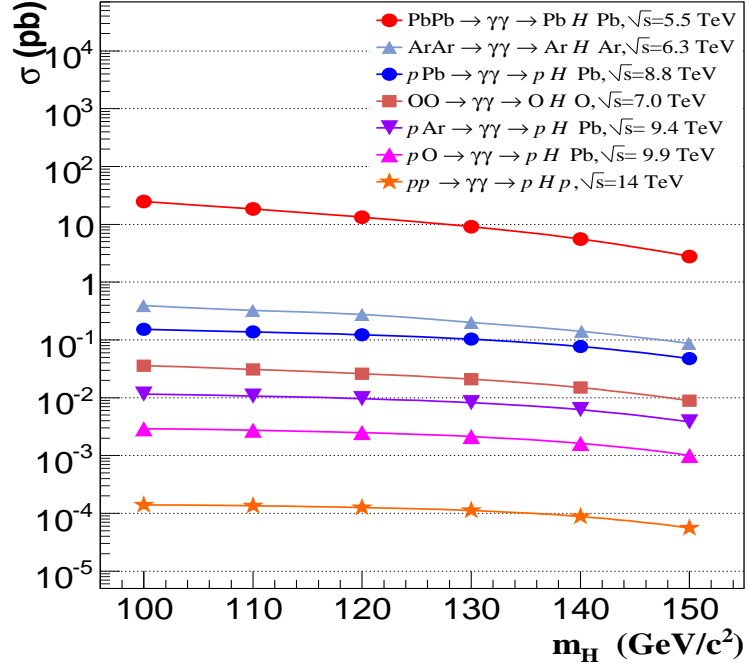}
\caption{Cross sections for the exclusive production of the SM Higgs versus its mass $m_H$ in elastic ultraperipheral
nucleus-nucleus, proton-nucleus and proton-proton collisions at the LHC (systems listed in Table~\ref{tab:1}).}
\label{fig:sigmaH_vs_mH}
\end{figure}

Table~\ref{tab:sigma_signal} lists the cross sections for Higgs ($m_H$~=~120~\GeVcc) production
in photon-photon collisions for the systems tabulated in Table~\ref{tab:1}. In Figure~\ref{fig:sigmaH_vs_mH} 
we show our predictions for the SM Higgs production cross sections as a function of $m_H$ for the same
systems. For $m_H$~=~120~\GeVcc, the cross sections span a range from 0.18~fb for \pp\ up to 18~pb 
for \PbPb. Compared to \pPb\ collisions, the ratios of the Higgs cross section 
between the different systems are roughly \pp\ : \pO\ : \pAr\ : \pPb\ : \PbPb\ = 1/900 : 1/50 : 1/13 : 1 : 100.
It is thus apparent that the large photon flux of the lead ion ($Z^2$) largely compensates for the higher projectile 
energies of the proton or light-ion beams, as well as the largest photon energies attainable with the smaller species.
However, when one takes into account the much smaller maximum luminosities at reach 
in the \PbPb\ running mode, such an advantage completely disappears (see Sect.~\ref{sec:rates_lumi}).\\

\begin{table}[htpb]
\begin{center}
\caption[]{Production cross sections for a SM Higgs boson with $m_H$~=~120 \GeVcc\ 
(total and for the $H\to\bbar$ decay) in elastic (\ABH) ultraperipheral collisions for the LHC 
colliding systems listed in Table~\ref{tab:1}. For \pp\ and \pA\ UPCs, we also quote the semielastic 
(\ABHX) cross sections (with the kinematical cuts $|y_{q'}|>2.5$ and $p_T^{q'}< 5$~\GeVc\ for the radiating quark).} 
\label{tab:sigma_signal}
\vspace{0.4cm}
\begin{tabular}{|l|c|c|c|c|c|} \hline\hline
System \hspace{0.1cm} & $\sqrt{s_{_{NN}}}$ & \multicolumn{2}{|c|}{\hspace{0.5cm} $\sigma(\gaga\to H)$ elastic (pb)\hspace{0.5cm}} 
& \multicolumn{2}{|c|}{\hspace{0.5cm}$\sigma(\gaga\to H)$ semielastic (pb)\hspace{0.5cm}} \\ 
& &  \multicolumn{2}{|c|}{\normalsize{[$m_H=120$~\GeVcc]}} & \multicolumn{2}{|c|}{\normalsize{[$m_H=120$~\GeVcc]}}   \\ 
& (TeV)  & \hspace{0.5cm} $H$ total \hspace{0.5cm} &  \hspace{0.5cm} $H\to\bbar$ \hspace{0.5cm} & \hspace{0.5cm} $H$ total \hspace{0.5cm} &  \hspace{0.5cm} $H\to\bbar$  \hspace{0.5cm}\\\hline
\pp  & 14 & 0.18$\cdot 10^{-3}$  & 0.13$\cdot 10^{-3}$ & 0.59$\cdot 10^{-3}$  & 0.45$\cdot 10^{-3}$ \\\hline
\pO  & 9.9 & 3.5$\cdot 10^{-3}$  & 2.5$\cdot 10^{-3}$ & 4.9$\cdot 10^{-3}$  & 3.5$\cdot 10^{-3}$ \\
\pAr & 9.4  & 1.3$\cdot 10^{-2}$  & 9.7$\cdot 10^{-3}$ & 1.7$\cdot 10^{-2}$  & 1.4$\cdot 10^{-2}$ \\
\pPb & 8.8 & 0.17  & 0.12   &  0.16  & 0.12 \\\hline
\OO  & 7.0 &  3.7$\cdot 10^{-2}$ & 2.6$\cdot 10^{-2}$ & --  & -- \\
\ArAr & 6.3 &  0.37 & 0.26 & --  & -- \\
\PbPb & 5.5 &  18   & 13 &  --  & -- \\\hline\hline 
\end{tabular}
\end{center}
\end{table}

The values we have obtained for the elastic exclusive Higgs cross sections for \pp\ and \AaAa\ (those for \pA\ are published in this work for the first time)
agree well with those found in the recent literature~\cite{Khoze:2001xm,Kryukov:2008zz} except for the \PbPb\ 
case where we overpredict the cross section by a factor two compared to~\cite{Kryukov:2008zz} due to the absence of the 
exact non-overlap condition in the convolution of fluxes as discussed previously. For \pp, the calculation of Khoze {\it et al.}~\cite{Khoze:2001xm} 
makes use of the standard formula for a narrow $\gaga$ resonance of spin $J$: 
$\sigma(\gamma\gamma \to H) \; = \; 8 \pi^2 (2J+1)\Gamma(H\to\gaga)/m_H^3 \: \delta(1 \: - \: m^2/m_H^2)$,  
which yields $\sigma(\gamma\gamma \to H) \; \simeq \; 0.12~{\rm fb}$, for $m_H = 120$~\GeVcc\ 
with width $\Gamma (H \rightarrow\gamma\gamma) \simeq 7.9$~keV/c$^2$~\cite{Djouadi:1997yw}.
This result takes into account an effective $\gaga$ luminosity 
of $1.1 \cdot 10^{-3}$ and a gap survival factor of $\hat{S}^2$~=~0.9, encoding the probability to produce 
fully exclusively the Higgs without any other hadronic activity from soft rescatterings between the protons.\\

The signal cross sections are enhanced by a factor of two by allowing for semielastic configurations where 
only the nucleus remains intact but the proton breaks apart after the photon emission from one of its quarks
(see right columns of Table~\ref{tab:sigma_signal}). This is the first time, as far as we can tell, that semielastic
UPC Higgs cross sections appear in the literature. As a cross-check, we have compared the semielastic 
cross sections for high-mass dileptons  in \pp\ UPCs obtained with our \madgraph\ prescription (see 
Sect.~\ref{sec:semiel}) with the results of~\cite{Bhattacharya:1995id} finding a good agreement. Note that in 
the \pp\ case the cross sections are multiplied by a factor of two since there are two possibilities (one for 
each proton) to emit the photon and fragment. For \pPb, we find a Higgs cross sections for semielastic 
$\gaga$ production, where the photon is radiated directly from the quarks of the proton, of 
$\sigma(\pPb \to \gamma\gamma \to X\, H\,Pb)\approx$~0.16~fb, 
i.e. very similar to that of coherent exclusive production. 
We do not quote the semielastic cross-sections for nucleus-nucleus UPCs since in these cases the photon 
flux, emitted either from their constituent protons or quarks, is much smaller than the coherent one.

\subsection{Background cross sections: $\gaga\to \bbar,\ccbar,\qqbar$}

The main background to the $H\to\bbar$ process is the continuum production of $\bbar$ and 
misidentified $\ccbar$ and $\qqbar$ ($q=u,d,s$) dijets. 
Table~\ref{tab:sigma_backgd} quotes the exclusive $\QQbar$ and $\qqbar$ cross sections in the range 
of invariant masses $m_{inv}$~=~100~--~140~\GeVcc\ relevant for a Higgs peak at 120~\GeVcc\ 
expected to be smeared by the $b$-jet reconstruction resolution (see Section~\ref{sec:Hsignific}). 
Without any kinematical cut the combined $\bbar$ continuum cross sections over $\Delta m$~=~40~\GeVcc\ is 
about 25 times larger (e.g. $\sigma_{\bbar}$~=~8~pb for \pPb) than the Higgs cross section at 120~\GeVcc\ (see 
Table~\ref{tab:sigma_signal}) for all systems. 

\begin{table}[htbp]
\begin{center} 
\caption[]{Production cross sections for exclusive production of 
$\bbar$, $\ccbar$ and \qqbar\ ($q\,=\,u,d,s$) with $m_{inv}$~=~100~--~140~\GeVcc\
and for $\ttbar$ (all masses) in elastic ($A \,B\,\xrightarrow{\gaga} A \,q\bar{q} \,B$) 
ultraperipheral collisions for the LHC colliding systems listed in Table~\ref{tab:1}. For \pp\ and 
\pA\ UPCs, we also quote the semielastic ($A \,B\,\xrightarrow{\gaga} X \,\bbar \,B$, with 
$|y_{q'}|>2.5$ and $p_T^{q'}< 5$~\GeVc\ for the photon-emitting quark) cross sections.} 
\label{tab:sigma_backgd}
\vspace{0.4cm}
\begin{tabular}{|l|c|c|c|c|c|c|} \hline\hline
System  \hspace{0.1cm}  & $\sqrt{s_{_{NN}}}$ & 
\multicolumn{2}{|c|}{$\sigma (\gaga\to \bbar)$~~(pb)} & 
$\sigma(\gaga\to \ccbar)$~~(pb) & $\sigma(\gaga\to \qqbar)$~~(pb) & $\sigma (\gaga\to \ttbar)$~~(pb) \\
&  
& \multicolumn{2}{|c|}{\small{[$m_{\bbar}$=100--140~\GeVcc]}} & 
\small{[$m_{\ccbar}$=100--140~\GeVcc]} & \small{[$m_{\qqbar}$=100--140~\GeVcc]} & \small{[all $m_{\ttbar}$]} \\
 & (TeV)    &  elastic  & semielastic  &  elastic &  elastic  &  elastic  \\\hline
\pp   & 14  &  3.4$\cdot 10^{-3}$ & 1.1$\cdot 10^{-2}$ &  7.9$\cdot 10^{-2}$ & 0.2 &  0.36$\cdot 10^{-3}$ \\\hline
\pO  & 9.9 &  6.8$\cdot 10^{-2}$ & 9.5$\cdot 10^{-2}$ &  1.6 & 3.9  &  2.7$\cdot 10^{-3}$ \\
\pAr & 9.4 &  0.27 & 0.36 & 6.1 & 15.8 & 8.1$\cdot 10^{-3}$ \\
\pPb & 8.8 &  3.4 & 4.5 &  78. & 200. &  6.2$\cdot 10^{-2}$ \\\hline
\OO  & 7.0 &  0.75 & --  & 17 & 39.  &  3.9$\cdot 10^{-2}$ \\
\ArAr & 6.3 & 7.6 & --  & 170 & 400. & 1.0$\cdot 10^{-2}$ \\
\PbPb & 5.5 & 420 &  -- & 9.4$\cdot 10^{3}$ & 2.5$\cdot 10^{4}$ & 1.8$\cdot 10^{-2}$  \\\hline\hline 
\end{tabular}
\end{center}
\end{table}


%
%
%
%
%
%
%
%

As we will see below, such irreducible background can be safely reduced with a few kinematics cuts. The $\ccbar$ 
($\qqbar$) cross sections over the same mass window are a factor of 600 (resp. 1600) larger than the Higgs signal 
but the probability of misidentifying both $c$-jets (resp. $q(\bar{q})$-jets) as $b$-jets is only of 0.25\% (resp. $2\cdot 10^{-4}$), 
and both backgrounds can be further reduced with the same selection criteria applied to remove the $\bbar$ continuum.
Note that, in agreement with our leading-order (LO) calculations for the signal, we do not consider the 
Next-to-Leading-Order (NLO) production of a heavy-quark dijet accompanied by an additional gluon radiated. 
This process is effectively  eliminated by our experimental requirement of two single jets in the event.\\

Table~\ref{tab:sigma_backgd} lists also the elastic top-antitop continuum inclusive cross section (for all $m_{\ttbar}$) 
which would be an interesting measurement in its own right, although as a potential $H\to\bbar$ background it can be easily 
removed given the presence of two extra $W$ decays ($\ttbar\to \bbar\,W^+\,W^-$) in those events.
These cross sections should be taken with a grain of salt since the very large $\gaga$ c.m. energies required
($m_{\ttbar}~\approx$~340~\GeVcc) are at the limit of applicability of our photon flux approximation.\\

We obtain a semielastic cross section of continuum $b$-quark dijets 
slightly larger (by a factor of $\sim$30\%) than the elastic ones. We do not quote the corresponding 
$\ccbar$ and $\qqbar$ semielastic cross sections which, for our signal over background studies 
(Sect.~\ref{sec:Hsignific}), we take also as a factor of 1.3 larger than the corresponding elastic values.

\subsection{Event rates and \pA\ luminosity considerations}
\label{sec:rates_lumi}

\subsubsection{\pA\ luminosities}

Taken at face value, the results 
of Fig.~\ref{fig:sigmaH_vs_mH} indicate that the 
$\gaga\to H$ cross section is maximum for lead-lead collisions and thus that 
this system should be the best suited to carry out such a measurement. Nevertheless, the \PbPb\ 
luminosity at the LHC is severely limited 
mostly due to two electromagnetic processes with huge cross sections that affect both Pb beams~\cite{Jowett:2008hb}:
(i) electron-positron production followed by $e^-$ capture by one of the nucleus (bound-free pair 
production, BFPP), Pb$^{82+}\;$Pb$^{82+}\xrightarrow{\gaga} $Pb$^{82+}\;$Pb$^{81+}\,e^+ $,
with a cross section of $\sigma_{\rm bfpp}$~=~280~b~\cite{Meier:2000ga}, and 
(ii) Coulomb dissociation of one or both nuclei due to mutual soft photon exchange(s), 
$^{208}$Pb$\;^{208}$Pb$\xrightarrow{\gamma} \,^{208}$Pb$\;^{207}$Pb$\,n $
with a cross section of $\sigma_{\rm emd}$~=~215~b~\cite{Pshenichnov:2001qd}. 
Both these processes create ions with a magnetic rigidity different than the nominal one 
for $^{208}Pb^{82+}$ ions, leading to beam losses and a reduction of the beam lifetime. In addition, the 
first process poses a danger of LHC magnet quenching due to the large amount of $Pb^{81+}$ 
ions straying from the nominal beam orbit, impinging on and heating the superconducting dipoles. 
Although other technical reasons limit the maximum luminosities attainable with ions (see below), 
BFPP effects effectively reduce the maximum luminosity in \PbPb\ collisions to
the $\mathcal{O}(10^{27}$~\Lunits) range, i.e. seven order of magnitude below the top
\pp\ LHC luminosity, $\mathcal{O}(10^{34}$~\Lunits).\\

In \pA\ collisions, the upper theoretical luminosity could be naively taken as the geometric mean of 
the maximum individual proton and ion beam luminosities, e.g. for \pPb,
$\Lumi_{pPb}^{max} =\sqrt{\Lumi_{Pb}^{max}\cdot\Lumi_p^{max}} =3\cdot 10^{30}$~\Lunits.
However, in principle the Pb-beam luminosities in \pPb\ could be significantly improved compared to \PbPb\ 
since the $\gaga \to e^+e^-$ cross section is $Z^2$~=~6700 times smaller and thus the quench-limit due
to BFPP could be naively raised by a $\mathcal{O}(7000)$ factor of up to
$\Lumi_{pPb}^{max} =\sqrt{\,6700 \cdot \Lumi_{Pb}^{max}\cdot \Lumi_p^{max}} = 2\cdot 10^{32}$~\Lunits. 
This is, however, only an idealistic estimate for several reasons.
The beam-beam luminosity in a generic $AB$ collision, $\Lumi_{AB}$, is given by the standard formula 
\begin{equation}
\Lumi_{AB} = \frac{{N_{b,A}\,N_{b,B} \;k_b \;f_0 \;\gamma }} {{4\pi \; \emittn \;\beta ^* }}\;F(\theta _c,\sigma^*,\sigma_z )\,,
\label{eq:lumidef}
\end{equation}
where $N_{b}$ is number of particles (protons or ions) per bunch in each beam, $k_b$ is the number of bunches per beam, 
$f_0=$~11.246~kHz is the revolution frequency (given by the LHC radius), 
$\gamma$ is the Lorentz factor (the geometric mean of the $\gamma$-factors of each beam for asymmetric systems), 
$\emittn=\sqrt{\gamma^2-1} \;\sigma_{x,y}^2 /\beta^*$ is the transverse normalized 
emittance related to the beam size $\sigma^*$, and $\beta^*$ the optical function 
at the interaction point (IP). 
$F(\theta_c,\sigma^*,\sigma_z )$  
is a  small reduction factor from the half-crossing angle, $\theta _c$, and bunch length $\sigma _z$,
which we neglect in this discussion. In Table~\ref{tab:pA_lumis} we list the beam parameters relevant for 
the \pA\ running mode~\cite{pAyellowreport,Jowett:2006au}.  We note that, as mentioned in the discussion 
of Table~\ref{tab:1}, the values quoted are for now only unofficial (but plausible) estimates.\\

\begin{table}[htbp]
\begin{center}
\caption{Basic beam parameters for protons and ions for 
\pA\ runs at the LHC: particles/bunch $N_b$, number of bunches $k_b$, 
normalized beam emittance $\emittn$, optics $\beta^*$, and associated luminosity $\Lumi_{AB}$.
Possible upgraded settings (see text) may lead to a factor of 60 improvement of the luminosities (last column).}
\vspace{0.2cm}
\begin{tabular}{|l|c|c|c|c|c|c|c|c|c|c|} \hline\hline
           & \multicolumn{9}{|c|}{nominal settings} & upgraded settings  \\ 
System & \multicolumn{2}{|c|}{$N_b$} & \multicolumn{2}{|c|}{$k_b$} & \multicolumn{2}{|c|}{$\emittn$ ($\mu$m)}
           & \multicolumn{2}{|c|}{$\beta^*$ (m)} & $\Lumi_{AB}$ & $\Lumi_{AB}$ \\  
           & proton & ion & proton & ion & proton & ion & proton & ion  &  (\Lunits) &  (\Lunits) \\\hline\hline
\pO\ (9.9 TeV)  & 1.15$\cdot 10^{10}$ & 1$\cdot 10^{9}$  & 2808 & 592 & 3.75 & 1.5 & 0.5 & 0.5 &  $2.7 \cdot 10^{30}$ & $1.6 \cdot 10^{32}$ \\\hline
\pAr\ (9.4 TeV)  & 1.15$\cdot 10^{10}$ & 5.5$\cdot 10^{8}$  & 2808 & 592 & 3.75& 1.5 & 0.5 & 0.5 & $1.5 \cdot 10^{30}$ & $1 \cdot 10^{32}$ \\\hline
\pPb\ (8.8 TeV)  & 1.15$\cdot 10^{10}$ & 7$\cdot 10^{7}$ & 2808 & 592 & 3.75  & 1.5 & 0.5 & 0.5 &  $1.5 \cdot 10^{29}$ & $1 \cdot 10^{31}$\\\hline\hline
\end{tabular}
\label{tab:pA_lumis}
\end{center}
\end{table}

Using these nominal beam parameters\footnote{For $k_b$ and $\emittn$ one uses the smallest of the proton 
or nucleus values.} and Eq.~(\ref{eq:lumidef}), we obtain the \pA\ luminosities listed 
in the before-last column of Table~\ref{tab:pA_lumis}. The obtained default \pPb\ luminosity is smaller
compared to the simple $\mathcal{O}(10^{30}-10^{32}$~\Lunits) estimates given above mainly because, 
conservatively, the proton intensity is reduced to 10\% of its standard value in \pp\ collisions~\cite{Jowett:2006au}. 
Note that the number of ions/bunch $N_b$ are significantly lower than the proton ones, mainly because of 
space charge effects in the SPS, and intra-beam scattering limits at injection in both SPS and LHC.\\

There are three potential paths to improve the \pA\ luminosity: (i) increase the proton
bunch intensity $N_b$ to its standard (10-times higher) value, (ii) increment the number of Pb 
bunches $k_b$ (which would be possible with the proposed new cryogenic collimators~\cite{Jowett:2009}) 
by a factor of 2--3, and (iii) carry out IP upgrades which should eventually allow factors of two
smaller $\beta^*$ at ATLAS and CMS. All such improvements are not unrealistic given that
the time-scale expected for a first proton-nucleus run at the LHC is at least 4--5 years after
the first \pp\ operation. By then, the knowledge of the collider and the upgrades related to
future LHC projects will be well advanced. The combined effect of such upgrades would optimistically allow 
one to increase the \pA\ luminosities by a factor of 60 (last column of Table~\ref{tab:pA_lumis}).\\

\subsubsection{Higgs event rates}

The expected number of Higgs bosons expected per year in ultraperipheral \pA\ collisions 
at the LHC can be obtained, from its production cross section (sum of elastic and inelastic channels) 
and the time-integrated luminosity, with the standard formula $N=\sigma_H\cdot\Lumi_{AB}\cdot\Delta t$.
The nominal LHC running time with protons (resp. ions) is 8 months (resp. 
1 month) which, with 50\% efficiency, corresponds to a run time of $\Delta t\approx$~10$^7$ 
(resp. 10$^6$)~s. Using the nominal luminosities quoted in Table~\ref{tab:pA_lumis},
we obtain the corresponding expected Higgs events for each one of the systems listed
in Table~\ref{tab:rates}.\\

\begin{table}[htpb]
\begin{center}
\caption[]{Expected Higgs boson ($m_H$~=~120 \GeVcc) production rates per year 
$N_{\rm Higgs}$ (in parenthesis those in the $\bbar$ channel) in elastic+semielastic ultraperipheral proton-proton, 
proton-nucleus and elastic nucleus-nucleus collisions at the LHC, for two run scenarios (see text). For each system, we quote the corresponding 
luminosity $\Lumi_{AB}$, the running time $\Delta t$, and the average number of overlapping pileup collisions, $\mean{N_{\rm pileup}}$.}
\label{tab:rates}
\vspace{0.2cm}
\begin{tabular}{|l|c|c|c|c|c|c|c|c|} \hline\hline
System & \multicolumn{4}{|c|}{nominal runs} &  \multicolumn{4}{|c|}{upgraded \pA\ scenario}  \\ 
& ${\cal L}_{AB}$ & $\Delta t$ & $\mean{N_{\rm pileup}}$ & $N_{\rm Higgs}$ & ${\cal L}_{AB}$ & $\Delta t$ & $\mean{N_{\rm pileup}}$ & $N_{\rm Higgs}$ \\  
            &  (\Lunits) & (s) & & total~~($H\to\bbar)$ & (\Lunits) & (s) &  & total~~($H\to\bbar)$  \\\hline
\pp\  (14 TeV) & 10$^{34}$ & 10$^{7}$ & 25 & 77.~~(55.) & 10$^{34}$ & 10$^{7}$  & 25 & 77.~~(55.) \\\hline
\pO\ (9.9 TeV)  & 2.7$\cdot$10$^{30}$ & 10$^{6}$ & 0.20 & 0.022~~(0.016) & 1.6$\cdot$10$^{32}$  & 10$^{7}$  & 3.9 & 13.~~(10.) \\
\pAr\ (9.4 TeV) & 1.5$\cdot$10$^{30}$ & 10$^{6}$ & 0.18  & 0.045~~(0.032) & 1$\cdot$10$^{32}$ & 10$^{7}$   & 3.6 & 30.~~(22.) \\
\pPb\ (8.8 TeV) & 1.5$\cdot$10$^{29}$ & 10$^{6}$  & 0.05 & 0.050~~(0.035) & 1$\cdot$10$^{31}$  & 10$^{7}$ & 1  & 34.~~(25.) \\\hline
\PbPb\ (5.5 TeV) & 5$\cdot$10$^{26}$  & 10$^{6}$  & 5$\cdot$10$^{-4}$ & 0.009~~(0.007) & 5$\cdot$10$^{26}$  & 10$^{7}$  & 5$\cdot$10$^{-4}$  & 0.15~~(0.1) \\\hline\hline
\end{tabular}
\end{center}
\end{table}

With the default settings and running times, the statistics are 
marginal for all systems involving nuclei. For the nominal runs, the possibility to carry out 
a measurement of the Higgs boson in photon-photon collisions at the LHC is virtually null, 
except maybe for \pp\ if one could work out a trigger -- e.g. using forward proton spectrometers~\cite{fp420} 
-- that can deal with the 20 inelastic proton-proton pileup collisions overlapping with the UPC event.
A straightforward way to increase the expected yields by a factor of 10 would be to 
dedicate a full LHC year (10$^7$~s) to a \pA\ run. This plus the upgraded luminosity settings
mentioned above would readily buy one a factor of 600 increase in the expected 
integrated luminosity (100~pb$^{-1}$) and, thus, of the number of Higgs counts per year (last column of~\ref{tab:rates}).
Unavoidably, the proposed improvements of the Pb-beam intensity and $\beta^\star$ optics imply an 
enhanced probability of having various collisions within the same bunch crossing. The occurrence of 
event pileup is particularly harmful in the case of UPCs since it eliminates the advantages given by the clean 
topologies of this type of collisions. The number of overlap collisions can be obtained from the product of the 
(inelastic\footnote{We do not care here about the ``harmless'' elastic interactions without particle production.}) 
reaction cross section  (last column of Table~\ref{tab:1}), 
the beam luminosity (last column of Table~\ref{tab:pA_lumis}) and the mean bunch distance: 
$\mean{N_{\rm pileup}}=\sigma_{AB} \times \Lumi_{AB} \times \mean{\Delta t_{bunch}}$,
the latter parameter depends on the revolution frequency of the beam 
and the number of bunches: $\mean{\Delta t_{bunch}}=1/(f_0\,k_b)$.
In the case of proton-nucleus, running with ions with the nominal bunch filling scheme, 
the crossing frequency is not well defined 
as not all encounters will occur at integer multiples of the 100-ns RF frequency~\cite{Jowett:2009}. 
For \pPb\ with $\Lumi$~=~1.5$\cdot$10$^{29}$~\Lunits, one has  $\mean{N_{\rm pileup}}\sim$~0.05, 
increasing to $\sim$1 if we consider the luminosity upgrades discussed above.
Clearly, the \pPb\ system provides the best combination of signal counting rates over pileup probability.


\section{Results II: $H\to\bbar$ measurement}
\label{sec:Hbbar}

\begin{figure}[htbp]
\centering
\includegraphics[width=0.60\columnwidth,height=6.5cm]{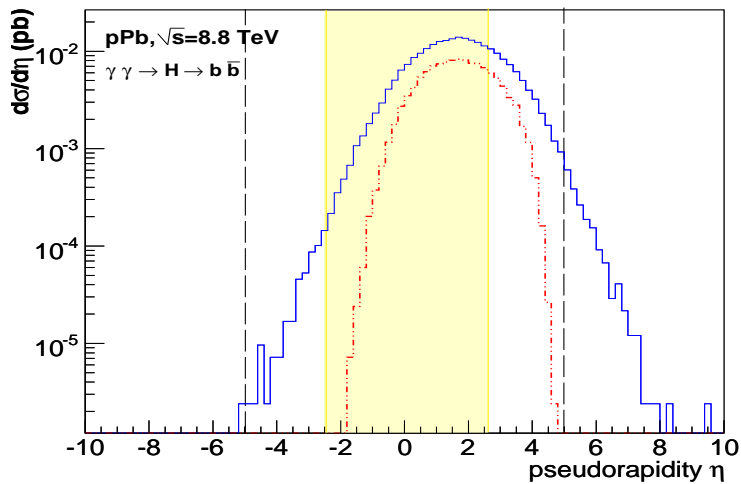}
\caption{
Rapidity-differential cross section for a Higgs boson ($m_H$~=~120~\GeVcc, dashed histogram) 
and its decay $\bbar$ jets (histogram) produced in \pPbH\ with $H\to\bbar$.  
The (shaded) yellow area indicates the acceptance of ATLAS/CMS detectors for $b$-jets.
The dashed lines indicate the full acceptance for all other type jets.}
\label{fig:ydistrib}
\end{figure}

In the last section of the paper we consider in detail the possibility to measure 
a 120~\GeVcc\ Higgs boson produced in ultraperipheral \pPb\ collisions at $\sqrtsnn$~=~8.8~TeV 
detected in the dominant $\bbar$ decay channel (BR~=~72\%) for this mass. 
Our analysis is entirely based on the ATLAS/CMS detectors (trackers, calorimeters), 
needed to reconstruct the $b$-jets and confirm the presence of a rapidity-gap, at central rapidities ($|\eta\,|<$~2.5).
No additional instrumentation is needed in principle except zero-degree calorimeters to help reduce possible diffractive 
proton-nucleus interactions (see below).
The generator-level rapidity ($y$) differential distributions for the $H\to\bbar$ signal (histogram) 
and decay $b$-jets (dashed histogram) 
are shown in Fig.~\ref{fig:ydistrib} (the semielastic distribution, not shown, is very similar). In our
calculations we take the direction of the proton-beam coming from negative rapidities.
We note that the $\gaga\to H$(120~\GeVcc) production is peaked forward, at $y=$~1.7 with an r.m.s 
of $\pm 1$ units of rapidity. On top of the $\Delta y_{\rm c.m.}$~=~0.47 shift due to the asymmetric 
$p$ and Pb beam energies, the proton EPA $\gamma$ spectrum is harder than the Pb one and boosts 
the production to even larger rapidities. The distribution of the two decay $b$-jets is centered at the 
parent Higgs rapidity but it is wider (r.m.s. of $\pm 1.4$).\\

The discussion presented hereafter will focus on the ATLAS and CMS experiments 
which feature $b$-jet reconstruction capabilities in the range needed to carry out the measurement
(shaded yellow area in Fig.~\ref{fig:ydistrib}).
The ALICE~\cite{alice-ppr1} acceptance for $b$-jet reconstruction is unfortunately limited to a 
narrow region $|\eta\,|<$~0.7 and there are luminosity limitations, $\mathcal{O}$(10$^{31}$~\Lunits), 
linked to the latency of the Time-Projection-Chamber. We do not consider here either the LHCb 
detector~\cite{lhcb}, although it features good $b$-jet identification capabilities at forward 
rapidities ($\eta\approx$~2--5) and covers a fraction of the Higgs decay acceptance in \pPb\ collisions 
(provided that the proton beam direction points in the same direction of the apparatus).\\

In order to obtain realistic estimates of the detectable number of Higgs in the $\bbar$ channel,
we have to consider the potential signal losses due to (i) the trigger efficiency (Sect.~\ref{sec:trigg}), 
(ii) the geometric acceptance of the detectors (Sect.~\ref{sec:bjet_accept}), 
(ii) the inefficiencies introduced by the finite-resolution and limitations of the experimental reconstruction 
(Sect.~\ref{sec:bjet_effic}), and (iii) the event selection cuts aiming at removing as much as 
possible backgrounds (Sects.~\ref{sec:bckgds}). We consider in detail these issues in the next subsections

\subsection{Triggering}
\label{sec:trigg}

There exist detailed trigger studies for ultraperipheral \AaAa\ collisions at the LHC~\cite{Baltz:2007kq,cms_hi_ptdr}. 
The main characteristics of (elastic) exclusive two-photon \pA\ events is the production of a 
single central particle accompanied by large rapidity gaps on both sides of it, and the survival 
of the interacting proton and nucleus. At variance with ultraperipheral \AaAa\ collisions, one cannot make use of the Zero Degree 
Calorimeters (ZDC)~\cite{zdcs} to tag the presence of a nucleus radiating a photon in ultraperipheral \pA\ collisions. 
In \AaAa\ UPCs, mutual Coulomb excitation of the incoming nuclei due to additional soft photon exchange(s) in the interaction,
produce forward neutrons from the decay of the excited nuclear Giant-Dipole-Resonance (GDR)~\cite{GDR}. 
The GDR excitation probability is proportional to the charge squared of the incoming projectile, which is a 
factor of $Z^2\approx$~6700 smaller for a proton than for a lead beam, and very few \pPb\ collisions will be
accompanied by forward neutron emission.\\

For most of the expected \pPb\ beam luminosities where no significant event pileup is expected
-- i.e. $\mean{N_{\rm pileup}}\lesssim$~1 proton-nucleus collision per bunch crossing -- one can easily record 
ultraperipheral Higgs events with virtually zero signal loss with a level-1 (L1) trigger based e.g. on 
(i) two back-to-back jets, with at least one of them with $p_T >$~40~\GeVc\, in the central detector ($|\eta\,|<$~2.5),
accompanied with an exclusivity condition given by (ii) a large rapidity gap $\Delta\eta\gtrsim$~2.5
without hadronic activity along the direction of the ion emitting the exchanged photon (whose energy 
is softer than the proton one). The rates of all signal and possible backgrounds with such signatures 
are well below a few Hz, easily allocatable without any prescale within the available L1 bandwidth.\\

In the semielastic case, only the nucleus survives and the rapidity gap is narrower along the proton direction 
due to the presence of forward hadrons from the fragmentation of the radiating quark and proton spectator 
partons. Yet, as discussed previously, our semielastic cross sections have been computed with the requirement 
of no activity within $|\eta\,|<$~2.5 other than the centrally produced system. In addition, since only the proton
dissociates the rapidity gap in the Pb direction is unpolluted by any hadronic activity.
The event topology of our $H\to\bbar$ semielastic events is basically identical to the pure elastic production and will 
identically pass the L1 trigger defined above.\\

For the maximum luminosities,  $\mathcal{O}(10^{31})$~\Lunits, considered in the ``upgraded \pA\ scenario''
one has to account for the possibility of 
one concurrent \pPb\ collision in the same bunch crossing as the $\gaga$ interaction (see Table~\ref{tab:rates}). 
In that case the rapidity-gap condition is not necessarily fulfilled since the exclusive event is overlapped by a normal 
(hadronic) proton-nucleus interaction. 
One can still use very-forward-proton detectors -- like TOTEM~\cite{totem} and ALFA~\cite{alpha} Roman Pots 
or the proposed FP420 spectrometer~\cite{fp420} -- to tag at the level-2 (L2) trigger, at least, the surviving proton in elastic $\gaga$ collisions.
The nucleus cannot be tagged similarly since its momentum transfer and energy loss are too small to leave the nominal LHC beam envelope. 
One-side tagging is a standard procedure for triggering on semielastic two-photon events in \pp\ collisions~\cite{louvain}.
One can still count on recording UPC events overlapping with a hadronic collision with 
dedicated level-1 high-$p_T$ $b$-jet triggers, 
plus a level-2 single-tagging of a leading proton, and separate offline the two interaction vertices 
(which is perfectly feasible since ATLAS and CMS can isolate the vertices of 25 overlapping collisions at the highest \pp\ 
luminosities). In the semielastic case, instead of the leading-proton one can tag the jet issuing from the fragmentation 
of the radiating quark in the forward calorimeters (FCAL in ATLAS, and HF/CASTOR in CMS).
All in all, such a scenario is less straightforward than the one considered for lower luminosities and would deserve 
a dedicated study beyond the scope of this exploratory paper. For the purpose of this study we will consider that L1 and L2 
triggers can be defined in \pPb\ collisions which are fully efficient with respect to the signal with counting rates in the Hz range that 
do not require any prescale factor.

\subsection{Experimental cuts}

Ideally, a complete MC simulation including parton showering and hadronization, full jet reconstruction and \geant-based 
detector response, would give fully realistic results for our study. This is, however, beyond the scope of this paper. 
We can, however, already obtain valid estimates of the feasibility of the measurement taking into account the 
known basic detector performances, and the kinematical properties of the signal and backgrounds at 
the generator-level.

\subsubsection{$b$-jet acceptance:} 
\label{sec:bjet_accept}

Full jet reconstruction in ATLAS and CMS is possible within $|\eta\,|<$~5 (or within $|\eta\,|<$~6.6  in CMS
if one includes the CASTOR calorimeter~\cite{castor}). Jet $b$-tagging requires however tens of micrometers 
vertex resolutions to identify the secondary vertex of the decay of the leading $B$-meson of the jet. Such capabilities 
are present only within the $|\eta\,|<$~2.5 central tracking coverage of the detectors\footnote{The combination 
of the CMS forward HF calorimeters and the TOTEM T1 trackers~\cite{totem} could potentially help to 
further extend the coverage for $b$-jets in ``particle-flow''-type analyses.} (see Fig.~\ref{fig:ydistrib}). A realistic 
ATLAS/CMS cut of the type
\begin{itemize}
\item both $b$-jets within $|\eta\,|<$~2.5,
\end{itemize}
has an acceptance of 55\% (i.e. it leads to a loss of 45\% of the signal) for both the elastic and semielastic
components. 
The acceptance for the dominant photon-photon $\qqbar$ continuum background (which has a wider distribution, 
see later) is fortunately smaller (23\%). 
One could in addition displace the vertex of the \pPb\ interaction point along the beam $z$-axis by up to about 
0.5~m by adjusting the optics in all experiments without much loss in the luminosity~\cite{Jowett:2009}. 
Such an offset in the proton direction would represent a gain of $\Delta \eta$~=~0.2 units of pseudorapidity 
for $b$-jets, and a corresponding increased geometric acceptance of 60\% for $H\to\bbar$.

\subsubsection{$b$-jet tagging efficiency:}
\label{sec:bjet_effic}

Standard $b$-jet reconstruction in ATLAS and CMS~\cite{atlas_tdr,cms_tdr} can be tuned to enhance 
either the $b$-tagging efficiency 
or the purity. On the one hand, due to the large charm $\gaga$ background -- with cross sections
$e_c^4/e_b^4$~=~16 times larger than $\bbar$ -- 
excellent $b$-tagging (i.e. increased purity) is required. On the other, one needs high tagging 
efficiency since the $\bbar$ reconstruction squares any single $b$-jet efficiency loss. Studying the ratio 
$(S/B)\propto \varepsilon_{b-tag}^2\cdot\sigma_{H\to\bbar}/[\varepsilon_{c-mistag}^2\cdot\sigma_{\ccbar} + \varepsilon_{q-mistag}^2\cdot\sigma_{\qqbar}]$ for various (correlated) 
(mis)tagging efficiencies from~\cite{atlas_tdr,cms_tdr}, 
leads us to choose a working point of 70\% $b$-jet tagging efficiency and a mis-tag rate of 5\% for $c$-quarks 
and 1.5\% for light-quark and gluon jets. The $c$-quark mistagging factor is a bit optimistic but 
it is likely at reach with multivariate-type analyses~\cite{Saout} after a few years of experience with previous \pp\ data.
Jet tagging efficiencies are slightly worst in the rapidities beyond $|\eta\,|\approx$~1.5 (for both 
signal and background jets), 
however the low particle multiplicities in two-photon events, likely compensate for such performance losses. 
In our analysis, we have therefore considered the following $\eta$-independent $b$-jet reconstruction
performances:
\begin{itemize}
\item $b$-jet tagging efficiency: 70\% for a single $b$-jet,
\item $b$-jet mistagging probabilities: 5\% for a $c$-quark, and 1.5\% for a light-quark,
\end{itemize}
which, for our {\it double} $b$-jet events of interest, lead to a $\sim$50\% efficiency for the signal and a total reduction 
of the $\ccbar$ and $\qqbar$ continuum backgrounds by factors of $\sim$400 and $\sim$4$\cdot 10^5$ 
respectively. 

\subsection{Background rejection}
\label{sec:bckgds}

Our ultimate goal is to have a number of Higgs events collected and a signal over background (S/B)
which is significant enough to observe  the $H\to\bbar$ channel at least at the $S/\sqrt{B} = 3\sigma$ level. We 
discuss in this section the procedure to remove in an offline analysis any remaining heavy-quark 
background that could have passed the triggers and experimental cuts discussed above. 

\subsubsection{Hadronic background:}
\label{sec:bckgd_had}

There may be peripheral (but still hadronic) grazing \pPb\ collisions with heavy-quark dijet production
and wide $\eta$ ranges without hadronic production (above experimental thresholds) which potentially pass 
our trigger selection criteria. 
The very low particle multiplicity and the very small transverse momentum of the centrally produced system 
($p_T\approx$~0 at leading-order but smeared by the experimental resolution) expected for the very low 
virtuality of the exchanged photons in UPCs can be successfully used to distinguish them from standard peripheral 
hadronic interactions:
\begin{itemize}
\item Jet multiplicity, $N_{jet}$~=~2: Requiring just two $b$-jets within $|\eta\,|<$~2.5 
selects with 100\% efficiency ``clean'' exclusive events and removes almost completely any 
hadronic interaction. This eliminates also possible genuine $\gaga\to H$ events where an additional 
jet from hard gluon emission is emitted, but this is consistent with the tree-level cross sections presented in the paper.
\item Exclusivity: Absence of hadronic activity (above detector backgrounds) outside the 
reconstructed jets, within $|\eta|\,<$~2.5 for tracks and neutral particles. For neutrals, one can 
extend the rapidity gap condition in the Pb direction up to $|\eta\,|<$~5 (or $|\eta\,|<$~6.6 
if one includes the CASTOR calorimeter in CMS), while still saving the semielastic signal.
\item Very low transverse momentum of the dijet system: $p_T^{pair}\lesssim$~5~\GeVc. Any photon-photon 
central system is expected to be produced almost at rest. 
Selecting events whose net $p_T$ is below a few \GeVc, to account for the experimental reconstruction
of a pair of two $b$-jets, eliminates any hadronic collision and still saves the semielastic Higgs component 
(with $p_T^H<$~5~\GeVc, see Sect.~\ref{sec:semiel}).
\end{itemize}
The application, at the offline analysis level, of such cuts removes virtually any remaining 
peripheral hadronic \pPb\ which could have been recorded at the trigger-level, with zero loss of the 
elastic and semielastic Higgs signals. In addition, if needed, one can also take into account the fact 
that the hadronic \pPb\ production of $\bbar$ jets is peaked at $y\approx$~0.47 (see 
Sect.~\ref{sec:sigma_calc}) whereas the two-photon fusion events are mostly centered 
at $y\approx$~1.7 (Fig.~\ref{fig:ydistrib}).

\subsubsection{Diffractive and photoproduction backgrounds:}
\label{sec:bckgd_diffract}

The experimental signatures of central diffractive (Pomeron-Pomeron or Pomeron-photon) interactions -- 
exclusive central object and two rapidity gaps -- are very similar to two-photon fusion processes~\cite{Engel:1996aa}. 
Central exclusive production of $\bbar$ jets above $p_T\approx$~40~\GeVc\ ($\Pom\Pom\to\bbar$) is a typical
background for exclusive Higgs production ($\Pom\Pom\to H\to\bbar$) with cross sections of the same 
order as the Higgs signal itself~\cite{fp420}.
Likewise, photoproduction of high-$p_T$ heavy-quarks -- in photon-gluon fusion via a $t$- or $u$-channel 
$\qqbar$ pair ($\gamma g\to\bbar$)~\cite{Vidovic:1995sm,Strikman:2005yv} or photon-Pomeron~\cite{Strikman:2005yv,Goncalves:2006xi} 
processes ($\gamma \Pom\to\bbar$) -- have event topologies (survival of the nucleus, one rapidity gap) 
similar to semielastic $\gaga$ production, and cross sections in \pPb\ collisions of order of a few nb 
for $b$-jets above $p_T\approx$~40~\GeVc, the purely diffractive contribution being $\sim$10\% of it~\cite{Strikman:2005yv,RVogt}.\\

Though both type of events have cross sections two to three orders of magnitude larger than our signal and they
can potentially pass the trigger cuts and contaminate our signal sample, 
there are various features that separate $\gaga$-events from $\Pom$-induced and $\gamma$-gluon,$\gamma$-$\Pom$ events:
\begin{itemize}
\item Photon-induced interactions are less central (i.e. take place at larger impact parameters)
than Pomeron-induced ones and, thus, 
the corresponding gap survival probabilities for masses $\mathcal{O}(100$~\GeVcc$)$ 
are much larger. 
The gap survival factor for Higgs production in \pp\ collisions at the LHC is  $\hat{S}^2_{\gaga} = 0.9$ in photon-fusion 
compared to $\hat{S}^2_{\Pom\Pom} = 0.03$ in Pomeron-fusion~\cite{Khoze:2001xm}. In the case of 
proton-nucleus collisions, the situation is comparably much more favorable for electromagnetic Higgs production: 
$\hat{S}^2_{\gaga} = 0.85$ versus $\hat{S}^2_{\Pom\Pom} = 8\cdot 10^{-4}$~\cite{Levin:2008gi}.
\item Since the nucleus is a fragile object -- the binding energy of a nucleon is just 8~MeV -- even the softest 
Pomeron-mediated interactions will result in the emission of a few nucleons from the ion, detectable in 
the zero degree calorimeters. On the contrary, in purely electromagnetic \pA\ interactions both hadrons
remain intact after the interaction and no other forward particles are emitted.
\item The net $p_T$ of elastic $\gaga$ final states is zero at LO. In the semielastic case, the typical $p_T$ 
of the produced Higgs is peaked below 1~\GeVc. In both cases, the total net $p_T$ is thus smaller than for 
comparable $\gamma$-nucleus or $\gamma$-proton interactions and thus is also an effective tool for separating the 
two classes of interactions~\cite{Baltz:2009jk}.
\item Due to the larger nucleus flux, heavy-quark photoproduction in \pPb\ is dominated by collisions of 
photons emitted by the lead nucleus with a gluon ($\gamma_{Pb}\,g_{p}$) or Pomeron ($\gamma_{Pb}\,\Pom_{p}$) 
that carry a larger fraction of the 7-TeV proton beam energy. The produced particles will be thus more forward 
boosted than our signal which comes from $\gamma_{Pb}\,\gamma_{p}$ collisions.
\end{itemize}
In short, inclusive high-$p_T$ heavy-quark photon- and Pomeron-induced production will partially fill one or both
rapidity-gaps and/or be accompanied by zero-degree neutrons, and the purely exclusive production for both processes 
has a gap-survival probability below $10^{-3}$. Thus, already before any kinematics cuts, both backgrounds are 
(much) smaller than our signal.\\

High-$p_T$ heavy-quark production can also take place via single (or double) resolved processes
with very energetic photons which interact via their partonic content~\cite{Drees:1995wh}
in collisions where the proton and the nucleus come very close together. We disregard this contribution 
in this analysis for various reasons. First, although the effective two-photon luminosity has indeed a large 
tail (Fig.~\ref{fig:lumis}), the photon fluxes are exponentially decreasing and the contributions to the cross sections 
in the high-$W_{\gamma\gamma}$ region relevant for such resolved processes are very small. Second, the 
much more energetic photon flux of the proton could be potentially resolved but at the price of an interaction
so close (with an impact parameter $b$ likely smaller than $R_p+R_{Pb}$) that the nucleus would 
break apart after the interaction. Our requirement for an intact nucleus implicitly sets a limit on the probability
to resolve the photon. Last but not least, any potential resolved photon contribution, would not only 
contribute to the heavy-quark background but also to the Higgs signal itself~\cite{Doncheski:2001uh}.

\subsubsection{Photon-photon continuum backgrounds:}
\label{sec:bckgd_gaga}

\begin{figure}[htbp]
\centering
\includegraphics[width=0.62\columnwidth]{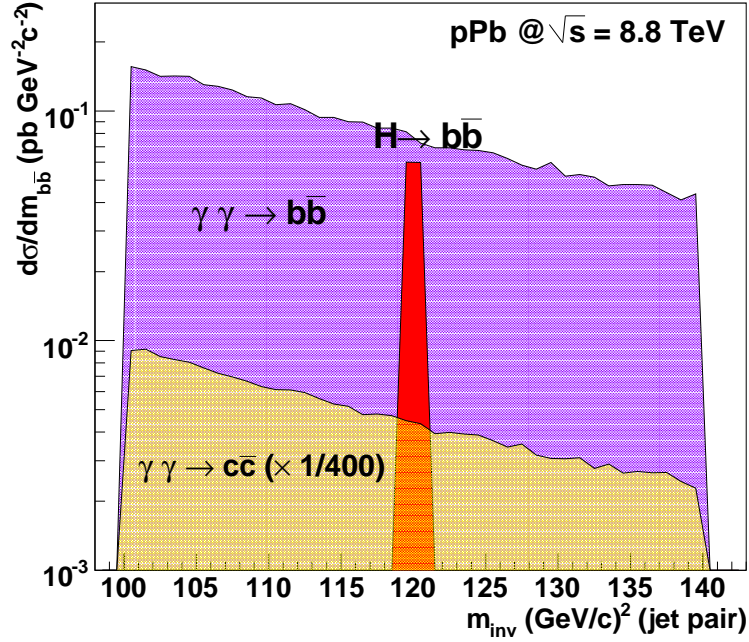}
\caption{Invariant mass distribution of pairs of $b$-jets from: (i) $H\to\bbar$ events and (ii) $\gaga\to\bbar$ 
and $\ccbar$ 
(misidentified with 0.25\% probability) continuum, in ultraperipheral \pPb\ collisions at 8.8 TeV. 
The Higgs peak is plotted with an (arbitrary) $\Delta m_H$~=~1~\GeVcc\ width.}
\label{fig:bjets_minv}
\end{figure}

The only physical backgrounds to $H\to\bbar$ in electromagnetic \pPb\ collisions are the (elastic and semielastic) 
exclusive $\gaga\to\QQbar,\qqbar$ processes with the cross-sections\footnote{The semielastic $\ccbar$ and $\qqbar$
continuum are taken to be 1.3 times the corresponding elastic cross sections as found in the $\bbar$ case.} quoted 
in Table~\ref{tab:sigma_backgd}. 
Other two-photon fusion processes, such as $\gaga\to\tau^+\tau^-$ or $\gaga\to \ttbar$ (see Table~\ref{tab:sigma_backgd}), 
have final-states (e.g. particle multiplicities) different than Higgs decay into two $b$-jets.
Although important (see Fig.~\ref{fig:bjets_minv}), the irreducible heavy-quark dijet backgrounds can be suppressed 
with various kinematics cuts.\\

\begin{figure}[htbp]
\centering
\includegraphics[width=0.49\columnwidth,height=6.6cm]{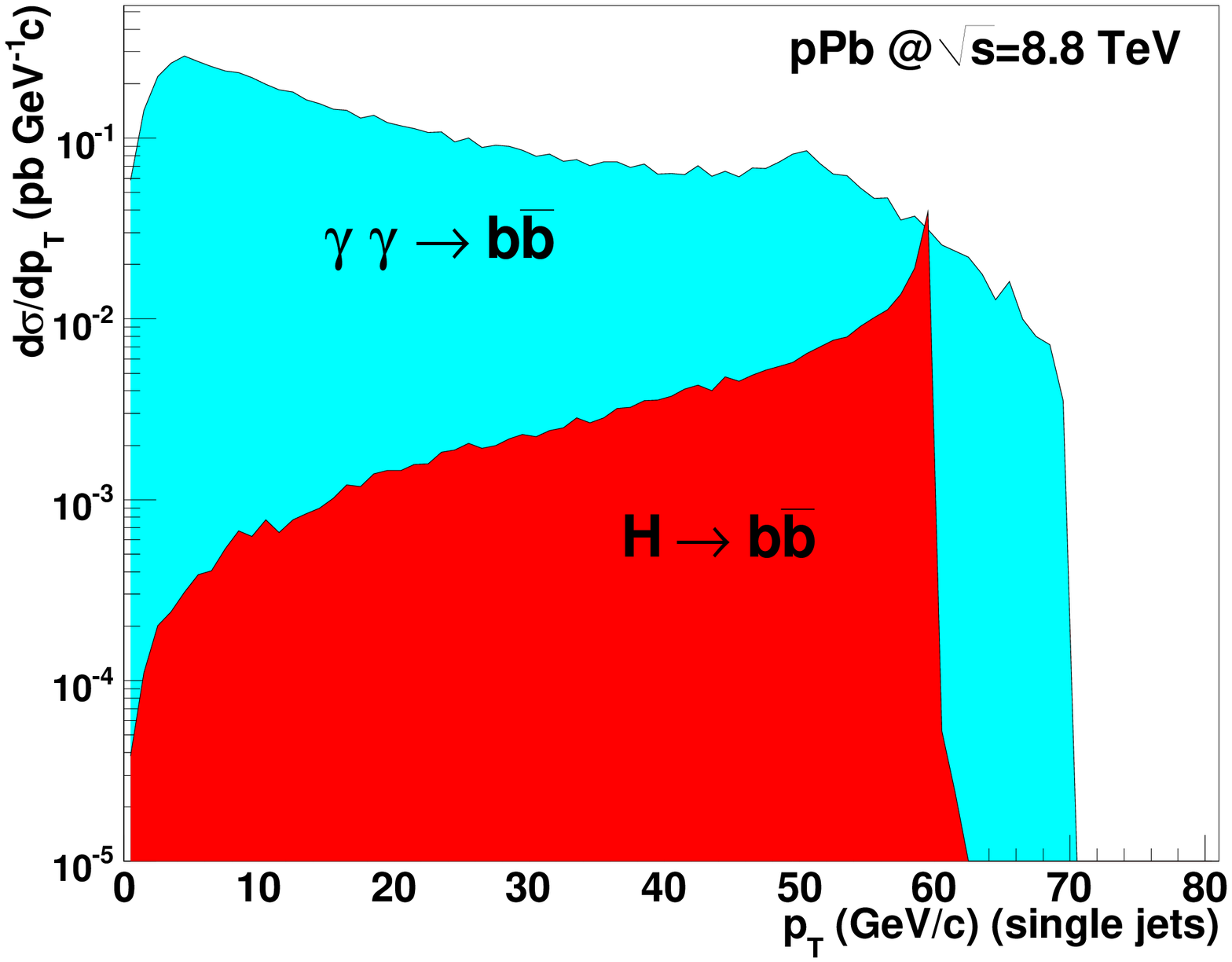}
\includegraphics[width=0.49\columnwidth,height=6.5cm]{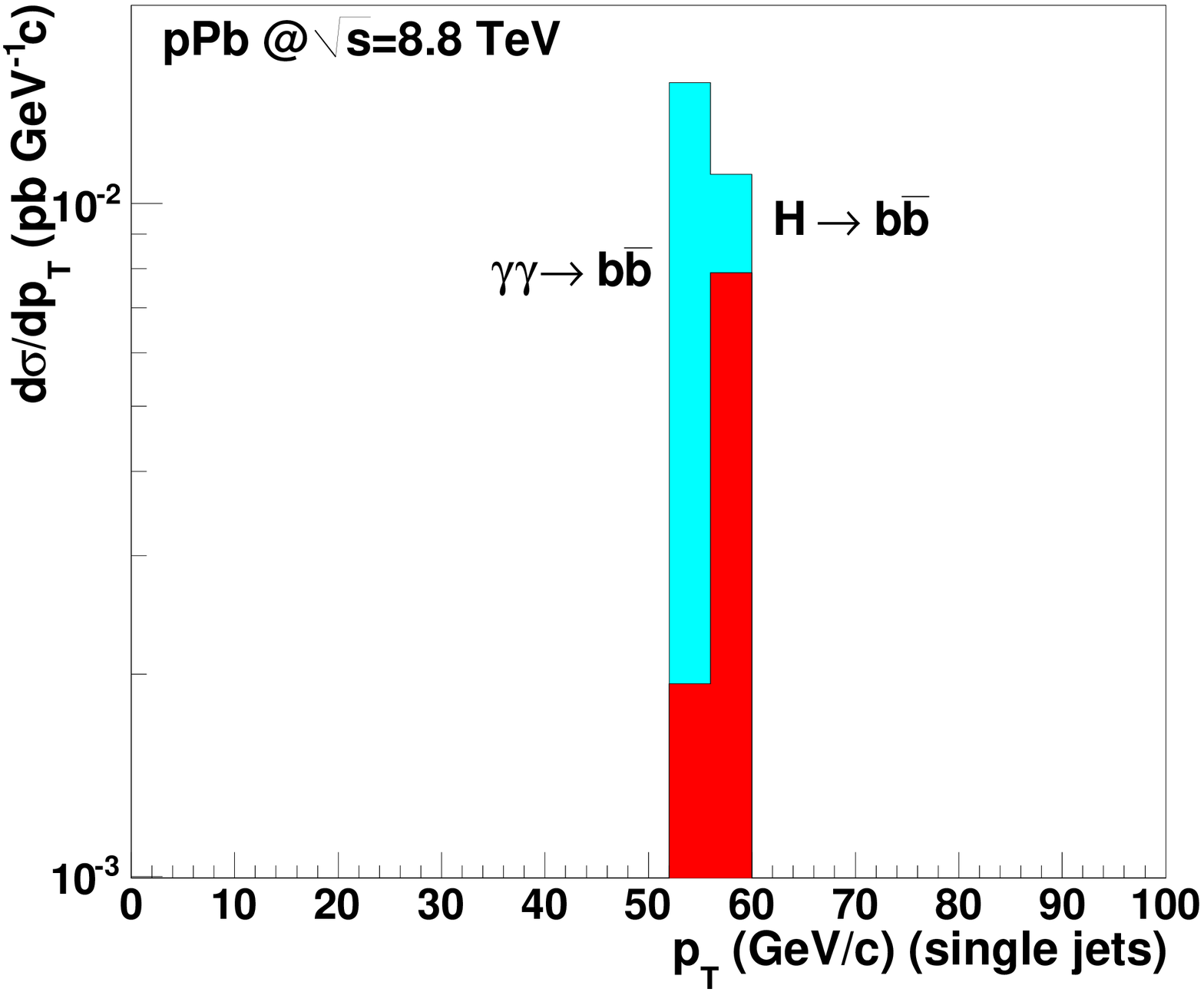}
\caption{Transverse momentum distribution of single $b$-jets from $H\to\bbar$  and 
$\gaga\to\bbar$ continuum events in ultraperipheral \pPb\ collisions at 8.8 TeV, before 
(pure MC-level, left) and after (right) applying all the experimental cuts discussed in the text.} 
\label{fig:bjets_pTdistrib}
\end{figure}

\begin{figure}[htbp]
\centering
\includegraphics[width=0.49\columnwidth,height=6.5cm]{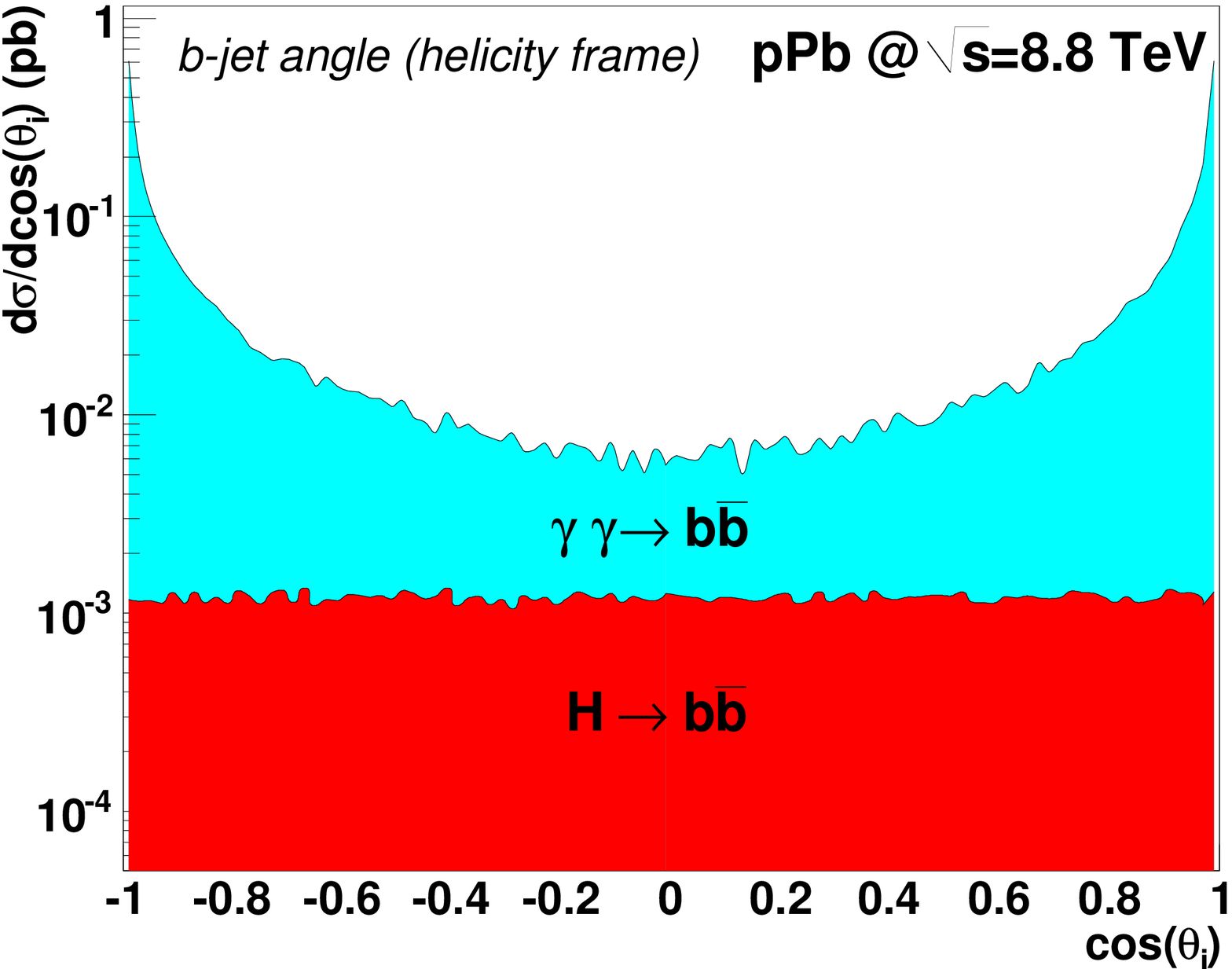}
\includegraphics[width=0.49\columnwidth,height=6.5cm]{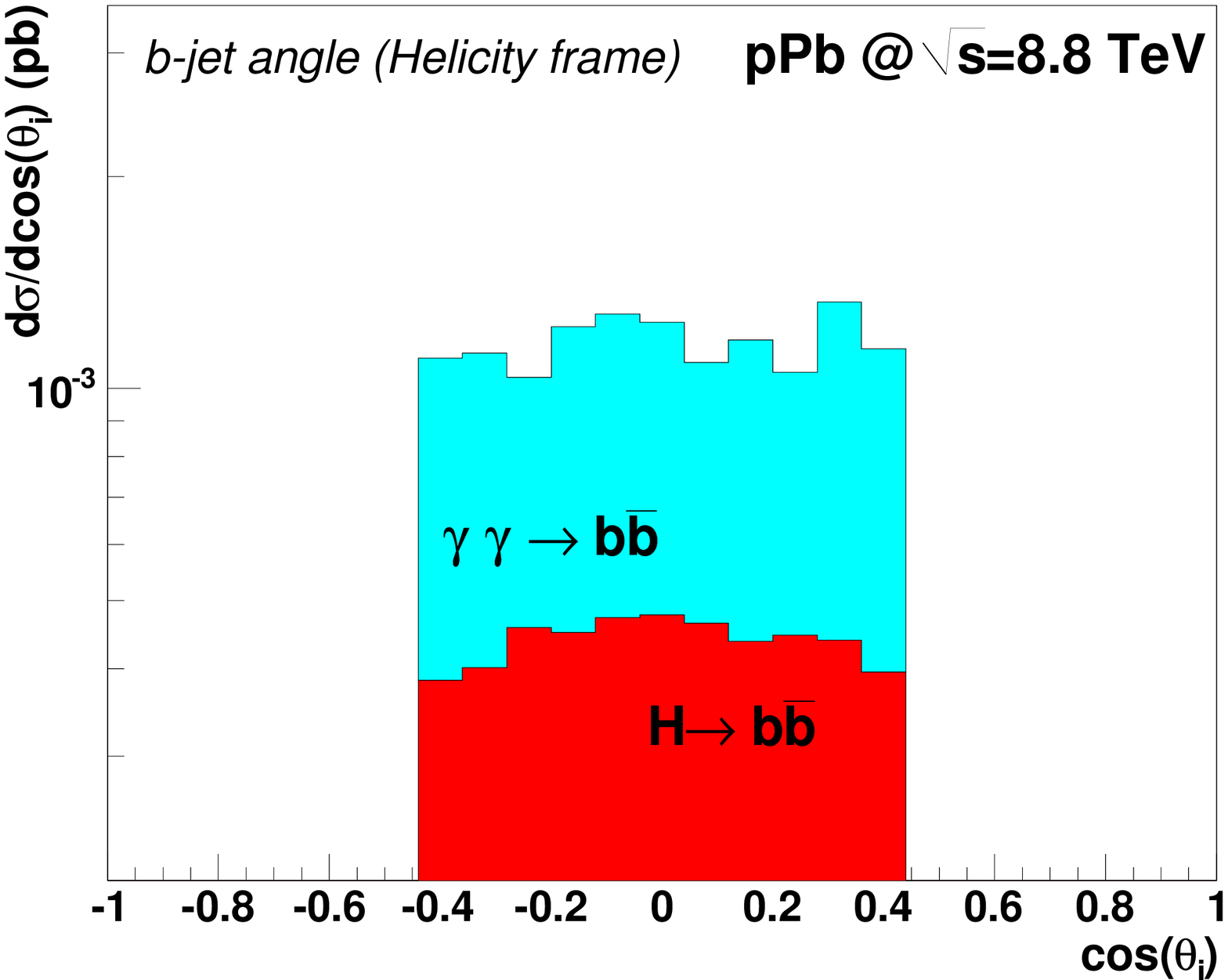}
\caption{Distribution of $\cos(\theta)$ for $b$-jets in the helicity frame from: (i) $\gaga\to\bbar$ continuum 
(top histograms) and (ii) $H\to\bbar$ signal (bottom histograms) in ultraperipheral \pPb\ collisions at 8.8 TeV, 
before (pure MC-level, left) and after (right) applying all the experimental cuts discussed in the text.}
\label{fig:bjets_costhetadistrib}
\end{figure}

First, the continuum can be reduced if we require the $b$-jet transverse momentum to be bigger than a significant 
fraction of the $\bbar$ invariant mass. In Fig.~\ref{fig:bjets_pTdistrib} (left), we compare the $p_T$ distribution of the 
signal and background before cuts. 
The background is dominated by one jet with low $p_T$ (but large $p_L$), whereas the signal 
peaks at $p_T\approx m_H/2$~=~60~\GeVc. Selecting events where there is at least one jet with 
$p_T$~=~52~--~60~\GeVc\ removes 96\% (97\%) of the $\bbar$ ($\ccbar$) background while killing 
only 49\% of the signal. 
Second, whereas the two Higgs decay $b$-jets are emitted isotropically, the continuum -- whose relevant Feynman 
diagrams have quarks propagating in the $t$- or $u$- channel -- is peaked in the forward and  backward directions 
(Fig.~\ref{fig:bjets_costhetadistrib}). Cutting on the angle $\theta$ between the $b$-jet (boosted to the rest 
frame of the pair\footnote{The $\bar{b}$-jet is at $\pi$ rads from it.}) and the direction of the pair (helicity 
frame\footnote{One can use alternatively the Gottfried-Jackson frame, which uses the direction of the beam. The results are unchanged.}),
removes an important fraction of the background. The $b$-jets from the continuum are clearly peaked at $|\cos\theta\,|\approx$~1,
i.e. emitted either roughly in the same direction as the pair or opposite to it.
With a rather strict $| \cos\theta\,| < 0.45$ cut, 82\% of the
continuum is suppressed for a 55\% signal loss.\\

Other cuts were tested, based e.g. on the rapidity separation $|y_b-y_{\bar b}|$ between jets, without
further background suppression power. The final set of cuts applied in our analysis is thus:
\begin{itemize}
\item Transverse momentum: at least one jet with $p_T^{jet}$ between $m_H/2.3$~=~52~\GeVc\ 
and the kinematical limit at $m_H/2$~=~60~\GeVc.
\item Acollinearity: $| \cos\theta\,| < 0.45$, where $\theta$ is the helicity-frame angle (between the $b$-jet, 
boosted to the rest frame of the pair, and the direction of the pair).
\item Mass window: Invariant mass of the $b$-jet pairs around the Higgs mass: $m_{\rm pair}$~=~100~--~140~\GeVcc\
(the range given by 2 or 3 times the width of experimental mass resolution).
\end{itemize}
We note that for a given event, these selection criteria (as well as the previous acceptance and $b$-tagging
efficiency cuts) do not necessarily factorize.

\subsection{Signal significance}
\label{sec:Hsignific}

The combined application of all experimental and background suppression cuts discussed in the previous
section leads to a  loss of about 80\% of the Higgs events and an important reduction of the $\bbar$, $\ccbar$ and 
$\qqbar$ continuum backgrounds by factors of 60, 5000 and 4$\cdot 10^{5}$ respectively. In one year 
(10$^7$~s) with the upgraded 10$^{31}$~\Lunits \pA\ luminosity scenario (i.e. integrating 100 pb$^{-1}$ of luminosity)
this corresponds 
to a collection of  $N$~=~5~Higgs and $N$~=~10~$\bbar$ continuum events within 100~--~140~\GeVcc\ 
with small $\ccbar$ ($N$~=~0.16) and negligible $\qqbar$ contributions. These values include both elastic and semielastic 
processes for the signal and backgrounds. The continuum decreases exponentially in this mass range whereas
the Higgs signal peaks at $m_H$~=~120~\GeVcc.\\

\begin{figure}[htbp]
\centering
\includegraphics[width=0.60\columnwidth,height=8.4cm]{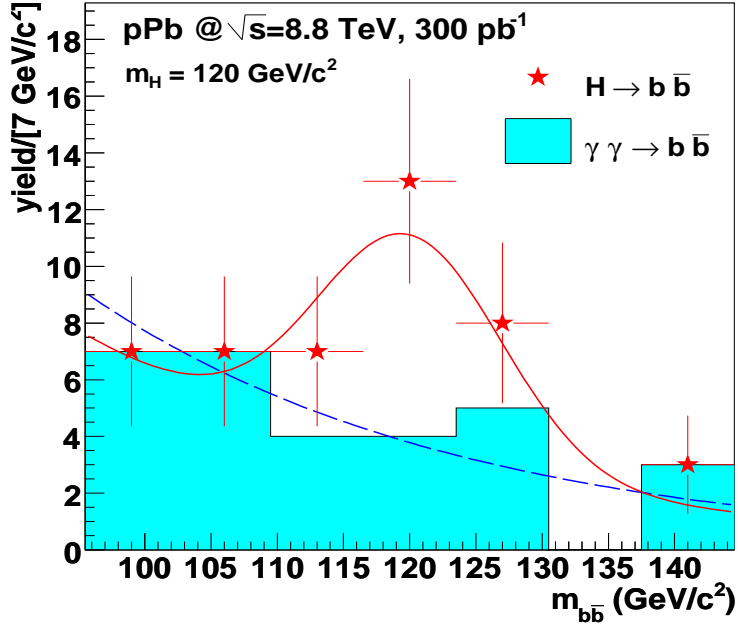}
\caption{Expected invariant mass distribution after 3-years of \pPb\ running at $\sqrtsnn$~=~8.8~TeV 
(300~pb$^{-1}$) for a $H\to\bbar$ signal 
and residual $\gaga\to\bbar$ continuum (histogram) 
after analysis cuts, fitted to a Gaussian+Exponential distribution (solid line). 
The data points are the sum of signal+background. The error bars are just statistical.
The dashed line shows an exponential fit of the continuum alone.} 
\label{fig:minv}
\end{figure}

In order to determine the significance of our $H\to\bbar$ signal, we generate  event samples consisting 
of the appropriate number of events after cuts in 3 years (3$\cdot$10$^7$~s) of data-taking,
i.e. 300~pb$^{-1}$ of integrated luminosity, with a fast Monte Carlo. We assume a $\bbar$ dijet 
invariant mass resolution of 7~\GeVcc. Such a value is beyond the current performances for $b$-jets 
in the range $p_T$~=~50~--~60~\GeVc, but may be achieved, in our underlying-event-free environment, 
with particle-flow reconstruction techniques and a good data-based knowledge of the 
$b$-jet $p_T$ resolution and energy scale after a few years of LHC running.
We then perform a log-likelihood fit to the pseudo-data with two curves: (i) one assuming that there is a signal+continuum 
in the region $100 \leq m_H \leq 140$~\GeVcc, and (ii) one assuming that there is just continuum. The significance, $S$, 
is then given by 
\begin{equation}
\mathcal{S} = \sqrt{\Delta \chi^2} 
\label{eq:signif}
\end{equation}
where $\Delta \chi^2$  is the difference of the $\chi^2$ of the fits for the signal and null hypotheses. We fit the 
peak region using a Gaussian fixed at $m_H$~=~120~\GeVcc\ (the Higgs mass, if any, will be already known by 
the time this measurement can be carried out) with width $\Delta m_{H}$~=~7~\GeVcc\ 
to match the assumed experimental mass resolution. We consider that the shape of the background will be known, since it 
can be measured with high statistics by simply removing the kinematics cuts applied to enhance the Higgs signal. 
According to our simulations, the exclusive continuum after cuts can be well reproduced by an exponential distribution
with inverse slope [28~\GeVcc]$^{-1}$, which we also fix in our fit. The normalisation of the signal and background are 
left as the only two free fit parameters. We repeat this method for 500 pseudo-data sets to obtain the average significance of the fits.
An example of a typical pseudo-data set and fits is shown in Fig.~\ref{fig:minv}. The error bars correspond to $\sqrt N$ 
statistical errors. For a total integrated luminosity of 300~pb$^{-1}$, we reach a signal-to-background ratio of 
$S_{H\to\bbar}/B_{\gaga\to\bbar}\approx 1.5$ and a statistical significance of $\mathcal{S} \approx 3$.\\

The main motivation for such a measurement is the unique observation of the SM $H\to \bbar$ decay which seems otherwise unaccessible
at the LHC. In addition, the observation of the $\gaga\to H$ process will provide an independent measurement of the Higgs-$\gamma$
coupling, likely measured before in the $H\to \gaga$ discovery channel. The $\gaga$-Higgs cross section is generated at the
one-loop level by all heavy charged particles ($W$ and top-quark in the SM) and it is thus sensitive to possible contributions of 
charged particles predicted in various extensions of the SM: e.g. chargino and top-squark loops in SUSY models,
and/or charged Higgs bosons in general 2-Higgs doublet models (2HDMs).
Last but not least, in the minimal SUSY extension of the SM -- which predicts three neutral Higgs bosons: the light CP-even $h$, 
the heavy CP-even $H$, and the CP-odd $A$ -- the properties of the of $h$ boson for large $A$ masses, are similar to the SM 
Higgs boson and it could be detected in the $\bbar$ decay mode as described in this work too. A dedicated study will 
be carried out to check the feasibility of such a measurement.



\section{Summary}


We have presented a detailed study of the exclusive production of the SM Higgs boson in electromagnetic (ultraperipheral)
proton-nucleus collisions at the LHC. We have evaluated the production cross sections and the 
corresponding yields via two-photon fusion processes in elastic (\pPbH) and semielastic (\pPbHX)
reactions, the latter being characterized by the breaking of the proton in the very forward
region. Such a measurement can be used to study, on the one hand, the $H-b$ coupling which is
otherwise not accessible to measurement at the LHC and, on the other, 
the photon-photon coupling to the Higgs, i.e. provides an independent check of the previously measured 
$H\to\gaga$ decay.\\

First, we have computed the Higgs boson cross sections in ultraperipheral \pp, \pA\ and \PbPb\ collisions
at LHC energies with the \madgraph\ Monte Carlo supplemented with equivalent photon spectra,
and demonstrated that \pPb\ collisions at $\sqrtsnn$~=~8.8~TeV give the best potential 
for such studies when realistic reachable luminosities and event pile-up issues are considered. 
In such a case, the total cross section of a Higgs boson of $m_H$ = 120~\GeVcc\  is about 0.15~pb for both
elastic and semielastic cases. The irreducible background due to the exclusive heavy-quark (and possibly 
misidentified light-quark) pair continuum, $\gaga\to\Qqbar,\qqbar$ has been also computed, 
along with the elastic $t \bar t$ production cross section, for electromagnetic \pAr, \pO\ and \pPb\ collisions.\\
 
In order to determine the feasibility of the $H\to b \bar b$ measurement, we have proceeded
to a detailed evaluation of the trigger setup needed, as well as the acceptances and efficiencies for 
the signal and continuum, and determined the best set of kinematical cuts needed to maximize the signal 
over background ratio. With reachable $b$-jet experimental reconstruction performances, we have found 
that a Higgs boson with $m_H$~=~120~\GeVcc\ could be observed in the $\bbar$ channel with a 
3$\sigma$-significance integrating 300~pb$^{-1}$ with an upgraded \pA\ luminosity of 10$^{31}$~\Lunits.\\
 
To conclude, while such a study will be rather demanding in terms of luminosity, we have shown that it
offers a unique complementary potential to the standard Higgs production mechanisms with regards to 
the study of its couplings to $b$-quarks and photons at the LHC. Both measurements are key to constrain 
the Standard Model and any of its possible extensions.


\section*{Acknowledgments}

\noindent
We thank Stan Brodsky, Albert de Roeck, Michael Peskin, and Mark Strikman for valuable 
discussions and suggestions on the paper. We also thank  Johan Alwall, Gerhard~Baur, Rikkert Frederix, 
John Jowett, Valery Khoze, Bernard Pire, Florian Schwennsen, and Ramona Vogt for useful communications. 
D.~d'E. acknowledges support by the 7th EU Framework Programme (contract FP7-ERG-2008-235071). 
This work is supported in part by the Belgian American Educational Foundation, the Francqui Foundation 
and the U.S. Department of Energy under contract number DE-AC02-76SF00515.



\end{document}